\documentclass[12pt,preprint]{aastex}

\usepackage{amsmath,amssymb,latexsym, graphicx}
\usepackage{epsfig}
\usepackage{booktabs}
\usepackage[dvips]{color}

\newcommand{\tn}[1]{\textnormal{#1}}

\newcommand{\kref}[1]{{\S} \ref{#1}}

\newcommand{\EBV}{E(B-V)}

\newcommand{\AV}{\tn{\emph{A}}_{\tn{V}}}
\newcommand{\RV}{\tn{R}_{\tn{V}}}
\newcommand{\NH}{\tn{N}_{\tn{H}}}

\newcommand{\MB}{\tn{M}_{\tn{B}}}

\slugcomment{Submitted to ApJ 25 May 2005, accepted for publication 16 December 2005, updated 22 December 2005}

\shorttitle{Dust in GRB afterglows}
\shortauthors{Kann et al.}

\begin{document}

\title{SIGNATURES OF EXTRAGALACTIC DUST IN PRE-\emph{SWIFT} GRB AFTERGLOWS}

\author{
D.~A.~Kann,\altaffilmark{1}
S.~Klose,\altaffilmark{1}
A.~Zeh\altaffilmark{1} 
}

\altaffiltext{1}{Th\"uringer Landessternwarte Tautenburg,  Sternwarte 5,
D--07778 Tautenburg, Germany}

\begin{abstract}
We present the results of a systematic analysis of gamma-ray burst afterglow
spectral energy distributions (SEDs) in the optical/near-infrared bands.  Our
input list includes the entire world sample of afterglows observed in the
pre-\emph{Swift} era by the end of 2004 that have sufficient publicly
available data. We apply various dust extinction models to fit the observed
SEDs (Milky Way, Large Magellanic Cloud and Small Magellanic Cloud) and derive
the corresponding intrinsic extinction in the GRB host galaxies and the
intrinsic spectral slopes of the afterglows. We then use these results to
explore the parameter space of the power-law index of the electron
distribution function and to derive the absolute magnitudes of the 
unextinguished afterglows.
\end{abstract}


\section{INTRODUCTION}

According to the most popular progenitor model of long-duration Gamma-Ray
Bursts (GRBs), the collapsar model \citep{Woosley1993}, a GRB is the result of
ultra-relativistic jets ejected by an accreting black hole formed by the
core-collapse of a massive star (most probably a Wolf-Rayet star). This
predicts a physical link between GRBs and supernovae (SNe) that has been
spectroscopically confirmed in four cases so far: XRF 020903
\citep{Soderberg2005}, GRB 021211/SN2002lt \citep{DellaValle2003}, GRB
030329/SN2003dh \citep[e.g.,][]{Hjorth030329, Stanek030329} and GRB
031203/SN2003lw \citep[e.g.,][Mazzali et al. 2005, in preparation]
{Malesani2004}. Further evidence comes from the observation of late-time bumps
in GRB afterglows that can be modelled very well by an underlying SN component
\citep{Bloom1999, ZKH}, and which  have led to the conclusion that in fact all
long-duration GRBs are  physically related to SN explosions \citep[][Paper
I]{ZKH}.  Furthermore, the collapsar model implies that the progenitors of
long-duration GRBs are associated with regions of high-mass star formation
\citep{Paczynski1998}, which might reveal themselves by a detectable
extinction in the GRB host galaxies along the lines of sight towards the
burster. This idea is further supported by the so-called dark bursts
\citep{Groot1998}, for which no optical afterglow has been discovered despite
rapid and deep searches in small error box regions. While most
none-discoveries are the result of too shallow searches and too large error
boxes \citep[e.g.,][]{Jakobsson2004b, Rol2005}, a small percentile remains
that require intrinsic extinction to dim the afterglow, e.g., GRB 970828
\citep{Groot1998, Djorgovski2001}, 990506 \citep{Taylor2000} and 020819
\citep{JakobssonIPR}, while others might have been intrinsically underluminous
\citep[for a discussion, see e.g.,][]{Fynbo2001b, Lazzati2002, Klose2003}.

Of particular interest within the dust extinction model is the  statistical
distribution of the amount of visual extinction in the GRB host galaxies, as
well as the nature of the corresponding dust. Given the fact that GRBs and
their afterglows can be observed up to high redshift, they offer the
possibility to get insight into the nature of cosmic dust when the universe
was much younger and galaxies were much less evolved. Furthermore, since the
optical properties of the dust grains trace the  environmental conditions in
the interstellar medium  \citep{FitzpatrickMassa1986, Draine2000, 
Bradley2005}, they are to some degree an indicator for the physical 
conditions to make a GRB progenitor.

The present paper is the third in a series of papers where we employ a large
database of photometry gathered from the literature to reanalyze all optical
afterglow light curves of GRBs in the pre-\emph{Swift} era in a consistent
manner to derive a homogenous sample, which is then used to study afterglow
properties in a statistical way. In Paper I and an update \citep{ZKH05}, the
properties of the supernovae underlying nearby afterglow light curves were
explored. In Paper II \citep{ZKK} the optical light curve parameters were
derived for a large (and basically complete) sample of afterglows observed by
the end of 2004, up to the launch of the \emph{Swift} mission. In this paper,
we extend this systematic analysis to the spectral energy distribution (SED)
of GRB afterglows  in the optical/near-infrared bands in order to search for
signals from extinction by dust in the GRB host galaxies.
 
In \kref{data}, we present the methods with which we analyzed the afterglow
SEDs. Among the 59 GRBs studied in Paper II, 30 had data quality sufficiently
high to be included in a sample for an investigation of the SEDs. We then
further reduce this sample to a Golden Sample of 19 GRB afterglows with
parameters derived from specific dust model fits. In \kref{RaD} we present the
results derived from our analysis and discuss our findings in the  context of
the standard fireball model. The fitting process allows us to derive the host
galaxy extinction $\AV$ along the line of sight and the intrinsic,
extinction-corrected spectral slope $\beta$ ($F_\nu\propto\nu^{-\beta}$) of
the afterglow light in the optical/near-infrared bands. The $\AV$-$\beta$
sample is then further analyzed statistically to derive conclusions about the
environment of GRBs and the dust properties of high-redshift galaxies. 

\section{DATA ANALYSIS}
\label{data}
\subsection{The fitting Procedure}
\label{analysis}

In Paper II we presented the analysis of a complete list of optical/NIR
afterglow light curves for which sufficient public data was available, up to
GRB 041006. One parameter derived in these light curve fits is the
normalization constant $m_k$, the magnitude of the afterglow at a certain
time, being one day after the GRB trigger in the case of a light curve fit
with a simple power-law ($F_\nu(t)\propto t^{-\alpha}$), or the jet break
time, $t_b$, in case of a fit with a smoothly broken power-law with a
pre-break decay slope $\alpha_1$ and a post-break decay slope $\alpha_2$
\citep {Beuermann1999}. All data used to fit these light curves is corrected
for Galactic extinction with $\EBV$ derived from the COBE maps \citep{SFD}
employing the Milky Way extinction curve of \cite{CCM}.  The contemporaneous
afterglow brightness in different photometric bands is then transformed into a
spectral energy distribution, using zero fluxes and median wavelengths taken
from the literature. The SED for GRB 030329 is derived in an alternate way,
see Appendix \ref{LitComp}. The fits use a Levenberg-Marquardt $\chi^2$
minimization algorithm. Unless stated otherwise, all errors in this paper are
at the $1\sigma$ confidence level.

Initially, we assume no extinction and the SEDs are fit with a simple
power-law, $F_\nu\propto\nu^{-\beta}$. In the following, we label a slope
derived from such a fit $\beta_0$. A steep $\beta_0$ combined with a
non-linear (curved) $F_\nu$ is then  indicative of extinction in the GRB host
galaxy.

In order to derive the visual extinction $\AV$ in the GRB host galaxy along
the line of sight, we transform the wavelengths of the SED into the host
galaxy frame following \citep{Fynbo2001}, using redshifts taken from the
compilation of \cite{FB2005}. In two cases redshift estimates are used. For
GRB 980519, we use a redshift of $z=1.5$, following \cite{Jaunsen2001} who
state that $z\geq1.5$ due to the lack of a supernova bump. For XRF 030723, we
varied the redshift and fit the SED for each redshift, and find the best
results in a redshift range of $z=0.3$ to 0.4. Thus, we adopt
$z=0.35$. \cite{FynboXRF030723} find tentative evidence for a low redshift
(although the host galaxy would be very faint), and \cite{ButlerXRF} find a
preferred $z=0.4$ from the Amati relation \citep{Amati2002}.  In both cases,
the SED is fit well by a power-law with an additional small amount of source
frame extinction. The influence of intergalactic Lyman absorption on the
photometry was accounted for by excluding SED data points lying beyond
$2.4\cdot 10^{15}$ Hz in the host frame from the fit\footnote{For GRB 030323
($z=3.37$), the V band also had to be excluded as the host galaxy is a DLA
with a very wide Lyman $\alpha$ absorption line \citep{Vreeswijk2004}.}. The
observed SED (corrected for Galactic extinction) is then modeled by the 
function
\begin{equation}
F_{\nu}=F_0\cdot\nu^{-\beta}\cdot e^{-\tau(\nu_{\tn{host}})}
\end{equation}
with
\begin{equation}
\tau(\nu_{host})=\frac{1}{1.086}\cdot \AV\cdot\eta(\nu_{host})\,.
\end{equation}
Here, $\beta$ is the intrinsic power-law slope of the SED, and $F_0$ a
normalization constant (we choose the unextinguished flux density at
$5\cdot10^{14}$ Hz in the host galaxy frame).  The function
$\eta(\nu_{host})=A_{\lambda _{\tn{host}}}/\AV$ is the extinction law assumed
for the interstellar medium (ISM) of the GRB host galaxy.  We call this
extinction source frame extinction. It encompasses local extinction close to
the site of the GRB and host galaxy extinction further away in case the
afterglow passes through a significant part of the host galaxy along the line
of sight. The extinction law for the Milky Way (MW), the Large Magellanic
Cloud (LMC), and the Small Magellanic Cloud (SMC) was taken from
\cite{Pei1992}. These three dust types differ strongly in the wavelength
region we examine (host frame UV/optical/NIR), especially toward shorter
wavelengths. The sequence MW, LMC, SMC is given by a decreasing strength of
the 2175 {\AA} bump, an increasing FUV extinction, and a decreasing reddening
per H atom \citep{Draine2000}. The former  implies the absence of very small
carbonaceous grains \citep{WD2001}, while  the latter is consistent  with the
trend observed for the metallicity in HII regions, with MW representing the
highest metallicity and SMC the lowest. Thus, SMC dust traces the dust
properties in low-metallicity environments. As the wavelengths of the SED are
shifted from the UV into the optical/NIR, a distinction between the three dust
types is easiest for GRBs at high redshift, whereas for nearby GRBs the three
extinction curves are almost identical in the considered wavelength regime.

Since the fit has three free parameters ($F_0$, $\beta$ and $\AV$), the
observed SED has to have at least four data points. The one exception is GRB
000131 (Appendix \ref{LitComp}), which we include because of
its very high redshift \citep[$z=4.5$,][]{Andersen2000}. The fitting process is
very sensitive to slight variations in the SED, especially if the $m_k$ data
points have small errors, so we did a careful check for each SED and removed
outliers. Furthermore, we added a systematic error of 0.03 magnitudes to each
data point of the SED (with the exception of those small number of cases where
$m_k$ is based on only one data point, as these almost always have much larger
errors anyway) to encompass uncertainties that may derive from the process of
data reduction. This results in a strong decrease of $\chi^2$ for the fits and
reduces the importance of this measure for discerning among different 
models for SED fits. 

\subsection{Selection of a Golden Sample}
\label{Golden}

Our final sample (Table \ref{tabALL}) comprises 30 afterglows out of the 59
bursts included in the sample of Paper II, with fits to the three dust models
for each afterglow. Figure \ref{SEDs} shows all 30 SEDs and the associated
best fits for each of the three dust models. This figure contains all fit
results that we obtain, including those with unphysical results,
just in order to show the failure of certain
extinction laws to reproduce the observational data.

Table \ref{tabALL} reveals that many of the fits are unsatisfactory. First of
all, some fits have very large error bars $\Delta\AV$ and $\Delta\beta$. Since
the SEDs are fit by allowing all three free parameters (incl. $F_0$) to vary
simultaneously, this creates a cross-sectional error, as the errors of $\beta$
and $\AV$ are correlated.  The steepest spectral slope ($\beta+\Delta\beta$)
results in the smallest extinction ($\AV-\Delta\AV$) and vice versa. Secondly,
many fits, especially when using the MW dust model, find \emph{negative}
extinction. This would imply bright emission features, for which we have no
evidence. Thirdly in some cases $\beta\leq0$ is found, while the standard
fireball model implies $\beta\geq0$ in the timespans (hours to days) and
wavelength bands that afterglows are typically observed in, and thus the
timespan where we derive $m_k$. These afterglow SEDs (e.g. GRB 971214) are
categorized by a relatively shallow unextinguished slope $\beta_0\approx0.5$ 
to 1 but a strong spectral curvature. Correction of this curvature results in
$\beta\leq0$. One solution would be dust with an increased FUV extinction, but
recent results on high redshift quasars \citep{Maiolino2004} do not support
this idea. We note that in these cases the data base is also sparse in 
photometric bands other than $R_C$.

To derive a more homogenous sample, which we call the Golden Sample, we employ
the following criteria:
\begin{enumerate}
\item The 1$\sigma$ error in $\beta$ ($\Delta\beta$) and the 1$\sigma$ error in
      $\AV$ ($\Delta\AV$) should both be $\leq0.5$.
\item $\AV+\Delta\AV\geq0$.
\item We do not consider GRBs where all fits (MW, LMC, and SMC) find $\AV<0$,
      even if the previous criterium is fulfilled. 
\item $\beta>0$ (although we do not reject cases with 
      $\beta-\Delta\beta\leq0$).
\item A known redshift.
\end{enumerate}
As there is no case where LMC dust is clearly preferred compared to SMC
and MW dust (Table \ref{tabALL}), in the following we remove all LMC dust 
cases from further consideration.

After applying these criteria, for the remaining 19 GRBs, in eight cases only
one dust model remains. For GRB 000131 no distinction can be made, as it is
not fit with extinction. The remaining ten GRBs give ambiguous results when
it comes to distinguishing between the two dust models. With the exception
of GRB 020813 and GRB 030328, where visual inspection of the fits shows that
the inclusion of SMC dust reproduces the observed SED better than MW dust, no
true distinction can be made at all. For these GRBs, we do not claim any dust
model preference, but for the sake of consistency, we choose the dust model
result which produces smaller errors in the derived parameters. This results
in 19 $\beta$-$\AV$ pairs which we designate our Golden Sample (Table
\ref{tabGold}).

\subsection{Highly extinguished GRBs outside our sample}
\label{Others}

There are some GRBs for which high visual source frame extinction has been
deduced in the literature but which are not included in Table \ref{tabALL}.
For GRB 980329, no redshift is known, although it is assumed to be high.
\cite{Jaunsen2003} derive a Bayesian photometric redshift of $z=3.6$.  The
very steep observed spectral slope (uncorrected for extinction) has been noted
by several authors. \cite{Yost2002} use their own data combined with data from
the literature and derive an $\AV\approx1$ for $z=3$, corresponding to
$\NH\approx2 \cdot10^{21}\tn{cm}^{-2}$ assuming a Galactic dust-to-gas
ratio. This is concurrent with our result $\AV=1.03\pm0.65$ for SMC dust and
under the assumption of a redshift of $z=3.6$. However, this result also has
$\beta\leq0$, implying that the true host extinction is lower than $\AV=1$.
Two afterglows which have redshifts but viable data in only two colors, GRB
990705 and GRB 000418, could not be fitted with the procedure outlined in
\S \ref{analysis}.  \cite{Masetti2000} find $\beta_0=2$, in agreement with our
result. While they do discuss reddening by intrinsic dust, their lack of a
redshift measurement does not allow them to find a definite result. The steep
initial decay suggests a wind environment with a cooling break blueward of the
optical bands. Using this case and SMC dust, we derive $\AV=1.15$. For GRB
000418, assuming a wind (ISM) environment with the cooling break blueward of
the optical bands and the new derived light curve parameters from Paper II we
fix $\beta$ via the $\alpha$-$\beta$ relations and  find $\AV=1.01$ ($\AV=
0.77$). The latter is slightly smaller than in \cite{Klose2000} (who used an ISM
environment), while \cite{Berger2001} find $\AV=0.4$ for LMC dust.  Finally,
there are indications of very high source frame extinction from the NIR
afterglows of GRB 030528 \citep{Rau2004} and GRB 040827 \citep{DeLuca040827},
but we were not able to perform any fits.

\section{RESULTS AND DISCUSSION}
\label{RaD}

\subsection{The prevalence of SMC-like dust in GRB host galaxies}
\label{SMCchapter}

In the selection of our Golden Sample, there were eight bursts where
application of our selection criteria yielded an unambiguous result.
SMC dust is by far the most preferred model, with seven out of the eight cases
(Table \ref{tabGold}), only GRB 970508 has evidence for MW dust. This result is
expected from studies of GRB host galaxies, which indicate that most hosts
are low-metallicity galaxies, as they are Lyman $\alpha$ emitters, blue, and
subluminous \citep[e.g.,][]{Fynbo2003, LeFloch2003, Jakobsson2005b}. Thus
we deduce that the ISM in these galaxies is probably metal-poor.

While the preference of SMC-like dust in GRB environments has already been
found by many groups for several bursts, e.g., GRB 000301C \citep{Jensen2001},
GRB 000926 \citep{Fynbo2001}, GRB 010222 \citep{Lee2001}, GRB 020813
\citep{Savaglio2004}, GRB 021004 \citep{Holland2003}, GRB 030226
\citep{Klose2004}, GRB 030429 \citep{Jakobsson2004}, XRF 030723
\citep{FynboXRF030723}, our study puts this conclusion on a statistical
basis. Given the sparse knowledge we have so far on the  dust properties in
cosmologically remote galaxies we note however that this finding does not
necessarily imply the requirement of low-metallicity environments for the
creation of GRB progenitors, as it might be indicated by SMC-like dust. The
fact that at least some afterglows SEDs are fitted best assuming MW dust
\citep[e.g., the spectral feature found in the spectrum of GRB 991216,] [which
is mirrored in our SED result, and the strong preference of MW dust for GRB
030227, although the large errors make this case unsure]{Vreeswijk2005} raises
doubt on such a general requirement. Even though one can not rule out the
possibility that in such cases the line of sight passes through foreground
material with different grain properties and ISM metallicity. In any case,
it is clear from this study that GRB afterglows can be used as a tool in order
to explore the properties of cosmic dust in the cosmologically remote
universe, in star-forming regions in particular and in galaxies in
general. But high quality photometric data are essential for building larger
samples.

\subsection{The extinction within GRB host galaxies}
\label{AVchapter}

Among the 19 GRBs in our Golden Sample (Table \ref{tabGold}), we find eleven
with $\AV-\Delta\AV>0$. We note that almost all of these cases are also those
with the highest quality data. Basically, this is not surprising, as more SED
data reduces errors. It is more speculative to state that, given high quality
SED data, there will always be small amounts of source frame extinction
detected in GRB afterglows. In eight of these cases, the detection is
significant at the 2$\sigma$ level or higher, it is significant at more than
3$\sigma$ in two cases and at 4$\sigma$ in one case.

The distribution of derived intrinsic extinctions is given in Figure
\ref{A_VHist}. A strong clustering toward low extinctions ($\AV\leq0.2$) is
evident. Zooming in with smaller bins shows that values between $\AV=0.1$ and
$\AV=0.2$ are preferred, with a non-error-weighted mean value of
$\overline{\AV}=0.21\pm0.04$. No afterglow is extinguished by more than 0.8
magnitudes. As a comparison, the prime example of a dark burst is plotted, GRB
970828. For this burst, \cite{Djorgovski2001} derive $\AV\gtrsim3.8$ by
interpolating between X-ray and radio data and comparing this with deep
optical observations that found no afterglow \cite{Groot1998}. The intervening
space is free of examples, creating a  potential Dark Burst Desert. On the
other hand, there are two afterglows in Table \ref{tabALL} that have mean
$\AV$ values lying in this Dark Burst Desert but are not included in the
Golden Sample due to the large errors of their fits (criterium (1) in
\S~\ref{Golden}). For GRB 010921, MW dust is preferred, as SMC dust finds
$\beta<0$. We derive $\AV=0.91\pm0.82$, in agreement with
\cite{Price2003}. While this is a positive detection, the error encompasses
most other bursts of the Golden Sample. For GRB 980703, no dust model
preference can be derived, but even for the SMC result with smaller
extinction, we find $\AV=1.32\pm0.59$.

How real is the non-existence of highly extinguished afterglows? A first
argument against the reality of the Dark Burst Desert is that our sample is
biased toward afterglows with low source frame extinction, as we require
viable photometric data in several wavebands and a known redshift. Intrinsic
extinction complicates the gathering of photometry (starting with the
discovery of the afterglow) and a spectroscopic redshift. In \kref{Others}, we
mentioned several probably highly extinguished afterglows that are not part of
our sample due to the sparsity of their photometric data. Highly extinguished
afterglows are not observed well enough to determine conclusively that they
are highly extinguished and not just intrinsically faint \citep{Klose2003}.

Figure \ref{A_Vtoz} shows the derived visual extinctions (Table \ref{tabGold})
plotted as a function of redshift $z$ \citep[values taken from][]{FB2005}. With
the exception of GRB 030429 the amount of extinction drops sharply toward high
redshifts. Most likely, this can be explained by an observational bias, at
least in the pre-\emph{Swift} era where only very few afterglows were already
localized within minutes after a burst. In particular, due to the wavelength
dependence of the dust opacity, the higher the redshift the more effective
dust can dim an afterglow in the optical bands. Furthermore, the redshift
measurement itself is biased toward bright and unextinguished afterglows, as
many redshifts are determined from absorption spectra taken when the afterglow
is much brighter than the underlying host galaxy. Finally, our data indicate
that, on average, low-$z$ GRBs have brighter afterglows (cf. Figure
\ref{BigPic1}), and thus have a greater chance of being detected through
significant extinction. Unfortunately, the chances of detecting a "Rosetta
Stone Dark Burst" at low redshift that is highly obscured by dust but still
bright enough to yield viable data and a redshift might be small if the GRB
frequency is coupled with the star-formation rate \citep[e.g.,][]{LR2002,
Firmani2004} and low metallicities \citep{MW1999}.

It is also visible that the the ability to discern between different dust
models is coupled to the redshift, as the strongest deviations (2175 {\AA}
bump, FUV extinction) appear in the UV region and the burst has to lie at a
certain redshift to shift these features into the optical bands.

\subsection{Host extinction versus star formation rate}
\label{submmchap}

Assuming that our database is not seriously affected by an observational bias
(dark bursts), Figure \ref{submm} shows that the standard GRB afterglow is
nearly unextinguished in its host galaxy. Since any long-lasting dust
destruction by the intense fireball radiation should represent itself in a
color evolution of the  afterglow \citep{Waxman2000}, which we do not find in
our data, we conclude that on average there is not much evidence for  dust in
GRB environments along the line of sight. Furthermore, any dust destruction
could only affect the dust in the immediate GRB environment, say, within 10 pc
around the burster \citep{Waxman2000, Draine2000, RhoadsFruchter2001,
DraineHao2002}. Dust at larger distances will still produce
extinction. Additionally, the \emph{Swift} satellite has made observations of
early GRB afterglows routine, and a major result is the almost complete
absence of reverse shock UV flashes \citep{Roming2005}. It seems that in most
cases the absence of  large amounts of dust is not due to dust destruction but
due to the fact that the standard GRB progenitor (seen pole-on along its
rotational axis) is not enshrouded by dust, globally and locally. Thus, if in
those cases where significant extinction is found the extinction is not just
produced in the immediate GRB environment (local extinction) but prevails
within the host galaxy (global extinction), we would expect that GRBs for
which we find significant optical extinction are located in very dusty,
submm-bright hosts.

In order to explore this possibility, we have taken from the literature all
submm data for GRB hosts with a flux density $F_{850}>0$ at the 1$\sigma$
level that are also in our main sample  of 30 afterglows \citep{Berger2003,
Tanvir2004} and calculated the corresponding star formation rate (SFR) via the
procedure developed by \cite{YC2002} \citep[cf.][]{Berger2003} to take the
different redshifts into account. Since the SFR is a direct measure of the
far-infrared luminosity of a galaxy \citep{K1998}, it traces its total amount
of radiating dust.  For GRB 980703, we conservatively used the result for SMC
dust, which has the lowest extinction, and took the SFR from
\cite{Berger2001b}. We added GRB 000418 ($\AV=1.01$, \kref{Others}),
arbitrarily assuming a 1~$\sigma$ error of 0.5 mag, and transformed the $\AV$
result for GRB 011211 (SMC dust) into an upper limit, as we find $\beta\leq0$,
making $\AV$ unsure.  The  resulting relation of SFR ($M_\odot$ per year)
versus $\AV$ is shown in Figure \ref{submm}. A trend is visible, a rising SFR
is coupled to a rising intrinsic visual extinction along the line of sight,
with the exception of GRB 010222. A linear fit to the data (excepting GRB
010222\footnote{ The most remarkable outlier in the potential SFR-$\AV$
relation is GRB 010222.  A Hubble Space Telescope image of the GRB 010222 host
galaxy reveals  the location of the GRB being offset by a small margin from
the center of the galaxy \citep{Fruchter2001, Galama2003}. We find a low
extinction value along the line of sight, whereas persistent submillimeter
flux indicates that this is a dusty starburst galaxy \citep{Frail2002}. The
discrepancy is resolved if the GRB happened toward the edge of the galaxy, but
our line of sight places it in front of the galaxy.})  gives
$SFR\;\mbox{$(M_\odot/$yr)}\;=220\pm130 + (500\pm180)\cdot\AV$.  We conclude
that, on average, GRB afterglows that show significant extinction along the
line of sight in their host are located in galaxies with a substantial star
formation rate and, hence, a globally acting extinction by large amounts of
dust. If this interpretation is correct, then the trend seen in the data
indicates that  the line-of-sight can pass through a significant extent of the
host galaxy.  Since the sample is still very small and the error bars on both
the SFR and the extinction $\AV$ are very large, it is clear that  more and
better data are required  in order to verify this result.

\subsection{The intrinsic spectral slope and the power-law
            index of the electron distribution function}
\label{betachapter}

Figure \ref{betaHist} displays the distribution of the intrinsic spectral
slopes, $\beta$, for the bursts from Table \ref{tabGold}.  The distribution is
broad, ranging from 0.2 to 1.2, and features a peak around $\beta=0.7$. We
find a non-error-weighted mean value of $\overline{\beta}=0.57\pm0.05$. This
result of a broad $\beta$ distribution of the intrinsic SED of the afterglows
is probably robust, since (1) our light curve fits of the individual
afterglows usually include many data points covering several days, so that we
are not much sensitive to individual measurement errors and (2) these fits do
not include the very early phase of an afterglow when its spectral properties
might develop much faster than at later times. In fact, with the exception of
the early afterglows of GRB 021004 and GRB 030329, we have never found clear
evidence for color variations of the genuine afterglow light in our data.

While it is not the goal of the present paper to find reliable explanations
for the observed width of the distribution of spectral slopes $\beta$ of the
afterglows in the optical/NIR bands, we note that the mean of this
distribution gives a value for the power-law index of the electron
distribution ($N(E)dE \propto E^{-p} dE$) of $p=2.4$, assuming a wind
environment and the cooling frequency blueward of the optical bands
\cite[][and references therin]{ZhangMeszaros2004} and the peak of the
histogram (Figure \ref{betaHist}). This is in general agreement with
theoretical predictions for ultra-relativistic shocks \citep{Kirk2000,
Achterberg2001}. The maximum values we find for $\beta$ (Figure
\ref{betaHist}) could then be explained by those afterglows which had
the cooling frequency redward of the optical bands during the entire
time span when they were observed, provided that
$p$ is a universal number. On the other hand, a universality of $p$ seems to
be difficult to reconcile with those afterglows which have $\beta<0.5$.
Assuming the standard afterglow models \citep{ZhangMeszaros2004}, these
afterglows require $p<2$, provided that the basic model assumptions are indeed
fulfilled in these cases. Note that in Paper II we also found afterglows that
require $p<2$ based on their light curve shapes alone.

The fact that some afterglows require $p<2$  has already been reported and in
detail explored by others \citep[cf. ][]{PK2000, PK2001, P2005}.  Based on our
SED fits three of 19 afterglows have $\beta\leq0.5$ within their 1$\sigma$
error bar (GRB 970508, GRB 991216 and GRB 030429) which implies $p<2$, while a
very shallow afterglow decay slope $\alpha_2$ brings GRB 990123, GRB 991216,
GRB 010222, GRB 030328 and GRB 041006 into this sample (Paper II). Obviously,
only a minority of afterglows require $p<2$.  It is not obvious why these
afterglows are specific in some sense.  The afterglow of GRB 991216 is one of
the brightest afterglow for redshifts $z\lesssim1$. On the other hand, the
afterglow of GRB 041006 is one of the less luminous afterglows
(\kref{redshifting}).  Thus, $p<2$ is not a question of luminosity.  Note that
both afterglows have only small intrinsic extinction ($\AV\approx0.1$), so
these low values for $\beta$ are not just an artifact of the fitting
process. For GRB 991216, the unextinguished slope is $\beta_0=0.54\pm0.03$,
for GRB 041006 it is $\beta_0=0.49\pm0.05$. We also note that all these
afterglow light curves are sampled fairly (GRB 990123) to very well (GRB
010222), implying that the finding of a flat spectral slope is not a question
of data quality either.

\subsection{Dust-to-gas ratios in GRB host galaxies}
\label{NH}

X-ray observations or the modeling of Lyman $\alpha$ absorption allow the
determination of the hydrogen column density $\NH$ along the line of sight
to the GRB in its host galaxy after correcting for the column density
in our Galaxy \citep{DL1990}. In Figure \ref{NHtoA_V} we present a sample of
afterglows for which we have derived $\AV$ and for which $\NH$ values are
reported in the literature (Table \ref{Xraydata}). This is an
update of the plot first presented in \cite{GalamaWijers2001} and expanded
in \cite{Stratta2004} \footnote{Note that we, unlike \cite{GalamaWijers2001}
and \cite{Stratta2004}, use a log-log plot to avoid crowding toward low
$\AV$, leading to a linear depiction of the dust model curves.}. The results
reinforce our findings presented in \kref{SMCchapter}. With two exceptions,
all points in the plot lie on the theoretical prediction for SMC dust or even
above it. The two exceptions are GRB 021004 and 030329, where only upper
limits for $\NH$ are given in the literature, leading to dust-to-gas ratios
higher than even for the Milky Way. On the other hand, both afterglow SEDs
are best fit with SMC dust (although the preference is only weak for GRB
030329), implying that the three dust models we use are not applicable in all
cases. Incidentally, GRB 021004 and GRB 030329 have the two best observed
afterglow light curves with the most pronounced substructure (Paper II). The
strongest outlier in Figure \ref{Xraydata} is GRB 990123, for which we find a
very low source frame extinction. The extremely bright UV flash of this burst
\citep{Akerlof1999} may have burned significant amounts of dust along the
line of sight \citep[cf.][]{Waxman2000, GalamaWijers2001, RhoadsFruchter2001,
PLF2003}, reducing the dust-to-gas ratio. No early multicolor data exists to
probe the color variations that are expected in the early light curve.

\subsection{The luminosity distribution of the afterglows}
\label{redshifting}

Knowledge of the intrinsic spectral slope of the afterglows allows us to
determine their luminosity distribution (Appendix \ref{Math}).

At first, in Figure \ref{BigPic1}, we show the $R_C$-band light curves of all
30 afterglows of the SED sample (Table \ref{tabALL}) plotted with
smoothed splines connecting the data points to guide the eye (for reasons
of clarity, error bars on the photometry have been omitted). The data have
been corrected for Galactic extinction and host contribution. For the
light curve of GRB 030329, the supernova contribution has also been
subtracted, using the data from \cite{ZKH05}. A very large spread of
magnitudes is seen, the range at one day after the trigger is 7.5
magnitudes between the afterglows of GRB 021211 and GRB 030329. Further
bright afterglows are those of GRB 991208 and GRB 991216, other faint
afterglows are GRB 040924, GRB 030227 and GRB 971214.

Figure \ref{BigPic2} displays the afterglow light curves of the Golden
Sample (Table \ref{tabGold}) after applying the cosmological $k$-correction
(the second term of eq. \ref{delta1}) (Table \ref{tabzCorr}) and after the
time shifting to a common redshift of $z_1=1$ (the first term of eq.
\ref{delta1}). In other words, Figure \ref{BigPic2} is a measure of the
absolute $R_C$ band magnitudes of the afterglows up to a constant. Compared
to Figure \ref{BigPic1} the magnitude range has decreased, now being 5.7
magnitudes at one day after the burst. Nine of the 16 afterglows that have
data one day after the burst lie in a range only two magnitudes wide,
approximately clustered around the afterglow of GRB 030329. In other
words, this afterglow is now seen to be quite typical. The two afterglows
above this range are those of GRB 021004 and GRB 991208 (assuming for the
latter burst an extrapolation of the decay with $\alpha\approx2.5$ to one
day after the burst in the host rest frame, Paper II). The six afterglows
beneath this range are GRB 030226, GRB 020405, GRB 030328, GRB 011121, GRB
041006 and GRB 040924. Of these six afterglows, only two (GRB 030226 and
GRB 030328) lie at $z>1$. GRB 021004 is the most luminous afterglow at all
times, although it is possible that the afterglow of GRB 991208 was
brighter at earlier times when it was not yet discovered
\citep{Castro-Tirado2001}\footnote{We note that both \cite{Nardini2005}
and \cite{LZ2005} have reached similar conclusions using host galaxy
extinction values derived from the literature.}.

Knowledge of the unextinguished light curves allows us to determine the
luminosity distribution of our afterglow sample.  Based on Fig.~\ref{BigPic2}
we derive the $R_C$ band magnitudes at one and four days (corresponding to
half a day and two days after the burst in the host frame  at $z$=1) and
transform them into absolute magnitudes $\MB$. We do not extrapolate the
afterglow light curves, except for GRB 030429 and GRB 020813, which both have
their final data points close to 4 days and are post-break. Thus, not all
light curves are included (e.g., GRB 991208 is not yet detected at one day,
GRB 030328 is not detected any more at four days if at $z=1$). The sample then
contains 16 GRBs at both one day and at four days, the results are given in
Table \ref{tabMB}.  Thereby, we computed the luminosity errors in a
conservative fashion.  Given that the errors of $\beta$ and $\AV$ are
correlated (the result $\AV+\Delta\AV$ is coupled to $\beta-\Delta\beta$ and
the other way around), we calculate three values of $\MB$ for the pairs
$(\beta,\;\AV)$, $(\beta+\Delta \beta,\;\AV-\Delta\AV)$, and
$(\beta-\Delta\beta,\;\AV+\Delta\AV)$. The pair
$(\beta+\Delta\beta,\;\AV-\Delta\AV)$ results in a lower luminosity, and the
pair $(\beta-\Delta\beta,\;\AV+\Delta\AV)$ in a higher luminosity. The
difference between these luminosities and the luminosity derived from the pair
($\beta,\;\AV$) is a conservative upper limit for  the error of $\MB$.  In
addition, we impose limits $\beta-\Delta\beta\geq0$ and
$\AV-\Delta\AV\geq0$. If $\beta-\Delta\beta<0$ or $\AV-\Delta\AV<0$, we set
$\beta=0$ or $\AV=0$, respectively, when computing $\MB$.

The resulting luminosity distributions are given in Figures \ref{MB1} and
\ref{MB4}. In addition, in order to search for a potential evolutionary effect
or an observational bias, we distinguish GRBs with $z<1.4$ and $z\geq1.4$, this
being the median of the redshift distribution of our sample. A bimodal
distribution of the afterglow luminosities is evident in Figures \ref{MB1z} and
\ref{MB4z}. Afterglows at $z<1.4$ tend to be less luminous, which might be
explainable by an observational bias: the chances to detect intrinsically
faint afterglows are higher for a lower $z$.

At one day after the GRB at $z=1$, the complete distribution has a two peaks,
the mean lying at $\overline{\MB}=-23.2\pm0.4$, with the most luminous
afterglow, GRB 021004, having $\MB=-25.58^{+0.04}_{-0.04}$, and the faintest,
GRB 040924, having $\MB=-19.84^{+0.17}_{-0.65}$. The means for the
distributions of the $z<1.4$ and $z\geq1.4$ GRBs are
$\overline{\MB}=-22.4\pm0.6$ and $\overline{\MB}=-24.1\pm0.5$,
respectively. The difference between the mean values is $1.7\pm0.7$ mag,
indicating a bimodality.

At four days after the burst at $z=1$, a single broad peak remains,
for the complete distribution, the mean value is $\overline{\MB}=-21.4
\pm0.4$. Now, the brightest afterglow (GRB 021004) has $\MB=-24.07^{+0.04}
_{-0.04}$ and the faintest afterglow (GRB 041006) has $\MB=-18.89^{+0.17}
_{-0.65}$. Once again, the bimodal distribution for nearby ($z<1.4$) and
distant ($z\geq1.4$) GRB afterglows is evident. The means for the
distributions of the $z<1.4$ and $z\geq1.4$ GRBs are $\overline{\MB}=
-21.2\pm0.5$ and $\overline{\MB}=-21.8\pm0.6$, respectively. The difference
between the mean values is $0.6\pm0.8$ mag, the significance of the
bimodality has been strongly reduced.

\cite{GB2005} analysed the X-ray afterglow light curves of GRBs and shifted
them to $z=1$ in an analog process, finding a bimodal flux distribution.
There are eight GRBs in their sample (GRB 970508, GRB 990123, GRB 991216,
GRB 000926, GRB 010222, GRB 011121, GRB 030226, and GRB 030329) that are
also among the afterglows we have plotted in Figure \ref{BigPic2}. GRB
990123, GRB 991216, GRB 000926, and GRB 010222 are in \emph{group I} of
\cite{GB2005}, while GRB 970508, GRB 011121, GRB 030226, and GRB 030329
are in their \emph{group II}. We note that for $z=1$, $t=1$ day lies at
the end of the initial plateau phase of the light curve of GRB 970508.
The afterglow brightens by 1.5 magnitudes shortly afterward. A comparison
reveals that while the two groups mix in the optical (the afterglow of
GRB 030329 being more luminous than those of GRB 990123 and GRB 010222
at one day after the burst at $z=1$), the mean absolute
magnitude of \emph{group I} afterglows is $\overline{\MB}=-24.2\pm
0.4$, while it is $\overline{\MB}=-22.7\pm0.5$ for \emph{group II}
afterglows, with the difference being $1.5\pm0.7$ mag. While this
finding is intriguing, the sample is too small to draw any conclusions.

Knowledge of the extinction corrected afterglow magnitudes allows us to
compare the luminosities of the afterglows at a common point in the evolution
of the jet, the jet break time. Ignoring any fine structure in the light
curves, for all bursts in our Golden Sample with a jet break (Paper II) we
computed the  apparent $R_C$-band afterglow magnitudes at the time of the jet
break from the Beuermann equation \citep{Beuermann1999}.  We included the fit
around the jet break of GRB 030329 but excluded GRB 021004, as the reality of
the late break we found in Paper II is unclear due to the many rebrightening
episodes. These magnitudes were then converted to luminosities and
normalized to the luminosity of the afterglow of GRB 990123 at the time of its
jet break.

We use these luminosity ratios to search for correlations between afterglow
luminosity and parameters of the prompt $\gamma$ emission of the GRBs. We take
the isotropic energy release $E_{\tn{iso}}$ and the source frame peak energy
$E_p$ from \cite{FB2005}, and the collimation corrected energies $E_\gamma$
from Paper II. We do not find any correlations between the afterglow
luminosity during the break time and these three parameters, with the absolute
value of the correlation coefficient being smaller than 0.25 in all cases.

\section{SUMMARY \& CONCLUSIONS}
\label{conclusions}

We have presented a sample of 30 GRB afterglow spectral energy distributions
in the optical/NIR bands which have been modeled with various dust extinction
curves (Milky Way, Large Magellanic Cloud and Small Magellanic Cloud) to
derive the source frame extinction, $\AV$, intrinsic to the host galaxies and
the spectral slope, $\beta$, of the afterglows unaffected by any dust
extinction. As all afterglows have been analyzed in a systematic way, the
results are fully comparable, making this sample unique in terms of both size
and consistency. For the further statistical study, we selected 19 afterglows,
our Golden Sample, which have physically reasonable results and small error
bars.

The preferred dust models we find (\kref{SMCchapter}) as well as the deduced
source frame dust-to-gas ratios (\kref{NH}) based on the inclusion of data
taken from the literature, both indicate that the majority of GRBs we have
investigated, covering the redshift range from 0.1 to 4.5, occur in
low-metallicity environments. The $\AV$ distribution that we have derived from
these data (\kref{AVchapter}) highlights a sparsity of strongly extinguished
afterglows, creating a Dark Burst Desert, even though it is unclear if the
preference of low extinctions is more than an observational and sample
selection bias. Our finding that most afterglows suffer from only low
extinction in their hosts could indicate that afterglows are usually not
obscured by dust close to the burster. One would then expect that the most
extinguished afterglows are in fact located in globally dusty hosts. Indeed,
we find weak evidence for  a correlation between the submm flux of GRB host
galaxies and  the source frame extinction $\AV$. Although the statistical
significance is low due to the small sample size and the large errors, this
finding calls for a more thorough investigation.

Knowledge of $\beta$ and $\AV$ allowed us then to correct the afterglow light
curves for intrinsic extinction and to derive the true luminosity distribution
of our afterglow sample at chosen times in the host galaxy frame
(\kref{redshifting}). We find that, on average, low-$z$ afterglows are less
luminous than high-$z$ afterglows. The most likely explanation we have at
hand for this finding is an observational bias against intrinsically faint
afterglows at high redshifts. A bimodal distribution found by \cite{GB2005} in
similarly corrected X-ray afterglows is not clearly seen in the optical
although, on average, GRBs with fainter X-ray afterglows also have fainter
optical afterglows. Unfortunately, the available sample size is still too
small to reach definite conclusions. A search for correlations between prompt
emission parameters and the luminosity of the optical afterglows at jet break
time has come up empty.

Since our sample is exclusively composed of GRBs from the pre-\emph{Swift}
era, a similar study in a few years time on \emph{Swift}-discovered GRB
afterglows will shed light on the Dark Burst Desert and the true afterglow
luminosity distribution by removing observational bias factors via rapid and
highly precise GRB localizations. Already, \emph{Swift} has lead to the
discovery of very faint afterglows \citep[e.g. GRB 050126, GRB 050607,]
[]{BergerSwift2005, Rhoads2005} including what may be the "darkest" burst
ever, GRB 050412 \citep{KosugiGCN, JakobssonIPR}. The recent discovery of
the first afterglows of short GRBs \citep{HjorthShort, FoxShort, CovinoShort,
BergerShort, SB051221} opens the possibility of finally making a comparison
of the environment of the two different classes of GRBs.

\acknowledgments We thank the anonymous referee for helpful comments that
improved this paper. D.A.K. and S.K. acknowledge financial support by DFG grant
Kl 766/13-2. A.Z. and S.K. acknowledge financial support by DFG grant Kl
766/11-1. We wish to thank S. Covino, J. Gorosabel, T. Kawabata, B. C. Lee,
K. Lindsay, E. Maiorano, N. Masetti, R. Sato, M. Uemura, P. M. Vreeswijk and
K. Wiersema for contributing unpublished or otherwise unavailable data to
the database, and S. Cortes (Clemson University) for reducing additional data.
D.A.K. wishes to thank N. Masetti and D. Malesani for enlightening
discussions. Furthermore, we wish to thank Scott Barthelmy, NASA for the
upkeep of the GCN Circulars and Jochen Greiner, Garching, for the "GRB Big
List".

\appendix

\section{Shifting the afterglows to a common redshift}
\label{Math}

In the following, we consider the proper afterglow, i.e., cleaned from any
underlying host   component as well as corrected for Galactic extinction and
extinction in the host galaxy. In cases where the late-time light curve is
dominated by a supernova component, the data was removed due to a lack of
knowledge concerning the supernova SED.  We start with the expression for the
flux density per unit frequency, $F(\nu, t)$,  of a time-dependent source that
is shifted to a redshift $z_1$, which was originally observed at a redshift
$z_0$ \citep{DDDR2002}:

\begin{eqnarray}
F (z_1; \nu, t) &=& \frac{1+z_1}{1+z_0}\
F \left(z_0; \nu \, \frac{1+z_1}{1+z_0}; t \, \frac{1+z_0}{1+z_1} \right)\;
\frac{d_L^2(z_0)}{d_L^2(z_1)}\ \exp\{\tau (\nu\,(1+z_0))\}\,.
\label{Dado}
\end{eqnarray}

\noindent 
Here, $t$ and $\nu$ are measured in the observers frame, $d_L$ is the
luminosity distance.  The exponential function corrects for a postulated
extinction in the host galaxy at redshift $z_0$, where $\tau$ is the optical
depth in the host at the observed frequency.

When shifting an afterglow to a redshift $z_1$, its apparent magnitude 
changes by an amount $\Delta m = m(z_1) - m(z_0)$, with
\begin{equation}
\Delta m    =-2.5\log\frac{\int_0^\infty S(\lambda)\, F(z_1;\nu,t)\, d\lambda}
                          {\int_0^\infty S(\lambda)\, F(z_0;\nu,t)\,
                          d\lambda}\,,
\label{dm}
\end{equation}
where $S(\lambda)$ is the wavelength-dependent filter response function for the
given photometric band. For an unabsorbed afterglow 

\begin{equation}
F(z_0; \nu,t)=\mbox{const.}\ f(t)\,\nu^{-\beta}\,.
\label{AG}
\end{equation}

\noindent We assume $\beta$ = const.
Since the light curve shape can be affected by a jet break and by 
rebrightening episodes (which are achromatic according to our present
data base), we do not specify $f(t)$. Using eq.~(\ref{AG}), it follows that 
in eq.~(\ref{Dado}) 

\begin{equation}
F \left(z_0; \nu \, \frac{1+z_1}{1+z_0}; t \, \frac{1+z_0}{1+z_1} \right)
= \mbox{const.}\ f(\tilde{t}) \, \nu^{-\beta}
\left(\frac{1+z_1}{1+z_0}\right)^{-\beta}\,,
\label{F1}
\end{equation}
\vspace*{2mm}
with $\tilde{t}= t \ \frac{1+z_0}{1+z_1}$. 
Any extinction correction is included in $F(z_1; \nu, t)$ in eq.~(\ref{Dado}).

After transforming flux density per unit frequency into flux density per unit
wavelength and inserting eqs.~(\ref{Dado}, \ref{AG},
\ref{F1})  into eq.~(\ref{dm}), we obtain
\begin{equation}
\Delta m    = - 2.5 \log \frac{f(\tilde{t})}{f(t)} - 2.5 \log\left(
\,\left[\frac{1+z_1}{1+z_0}\right]^{1-\beta}\
                \frac{d_L(z_0)^2}{d_L(z_1)^2}\,\frac{\int_0^\infty S(\lambda) 
\, \lambda^{\beta-2} 
                      \exp\{\tau(\lambda/(1+z_0))\}\, d\lambda}
                {\int_0^\infty S(\lambda) \, \lambda^{\beta-2} 
\,d\lambda}\right)
\label{delta1}
\end{equation}
In a logarithmic plot (apparent magnitude vs. log time) the first term in
eq.~(\ref{delta1}) represents a shift of the observed light curve  in time.
The second term in eq.~(\ref{delta1}) shifts the light curve along the
magnitude axis; the light curve shape is not affected.

In our calculations we assumed a flat universe with matter density $\Omega_M =
0.27$, cosmological constant $\Omega_\Lambda=0.73$, and Hubble Constant
$H_0=71$ km s$^{-1}$ Mpc$^{-1}$ \citep{Spergel2003}.  For $S(\lambda)$ we used
the filter functions of the VLT/FORS1 Bessel filters (see
http://www.eso.org/instruments/fors1/filters.html).

\section{Comparison with the literature}
\label{LitComp}

We have searched the literature for publications that also derive the source
frame extinction toward the gamma-ray-bursts listed in  Table~\ref{tabALL}. In
most cases, these results are in good agreement with our findings.

\emph{GRB 970508} To our knowledge, \cite{Reichart1998}  was the first who
discussed the extinction in a GRB host galaxy derived via a fit of the
observed SED in the optical bands.  Analysing collected data from the
literature, he finds for GRB 970508 $\AV=0.24^{+0.12}_{-0.08}$ (albeit for a
slightly higher $z$, having left the redshift as a free parameter), concurrent
with our result $\AV=0.38\pm0.11$ for the preferred MW dust extinction model.

\emph{GRB 971214} \cite{Halpern1998} find a strong spectral curvature in their
data and infer a large $\EBV\approx0.4$ for an assumed redshift of $z=2$. As a
fixed value for the spectral slope, they took the $V-I$ color of the afterglow
of GRB 970508.  This strong spectral curvature is mirrored in our SED fit
which finds a large extinction and $\beta\leq0$. Fits with a fixed $\beta$
derived from the $\alpha$-$\beta$ relations \citep{ZhangMeszaros2004} are
strongly rejected.

\emph{GRB 980519} We have found no references to intrinsic extinction in the
literature. This makes our preferred SMC dust result of $\AV=0.22\pm0.19$,
derived under the assumption of a redshift of $z=1.5$, a novel
result. 

\emph{GRB 980703} This burst has the highest extinction in our sample (Table
\ref{tabALL}). This result is not unprecedented, however. While finding
different values, several authors reported on a high extinction at different
times after the burst: \cite{Castro-Tirado1999} find $\AV=2.2$ 0.9 days
post-burst, \cite{Vreeswijk1999} find $\AV=1.5\pm0.11$ at 1.2 days, and
\cite{Bloom1998_2} derive $\AV=0.9\pm0.2$ at 5.3 days. Our achromatic fit
gives $\AV=1.93\pm0.91$ for the MW dust model, close to the result from
\cite{Castro-Tirado1999}, while SMC dust gives $\AV=1.32\pm0.59$. The light
curves are not sampled well enough to check if the discrepancies between the
various authors are the result of chromatic changes of the SED, a bumpy
structure, or a true decrease of line-of-sight extinction with time.

\emph{GRB 990123} \cite{Galama1999} find a negative extinction for this burst
and therefore fix it to zero, consistent with the low value we derive:
$\AV=0.04\pm0.05$ for the preferred SMC dust model. On the other hand,
\cite{Savaglio2003} derive a very high extinction of $\AV=1.1$ from metal
column abundances. Given the high quality of the optical photometry, we can
not reproduce such a high extinction.

\emph{GRB 990510} Both, \cite{Stanek1999} and \cite{Beuermann1999} find a
slight curvature in the SED. Removing the $B$-band data point,
\cite{Stanek1999} find $\beta=0.46\pm0.08$, comparable to our result,
$\beta=0.30\pm0.69$ for SMC dust.  Neither publication fits the SED with a
dust model to derive the host extinction.

\emph{GRB 991208} Almost all data on this afterglow (in $BVR_CI_C$) is from
\cite{Castro-Tirado2001}.  They do not discuss intrinsic extinction in the
host galaxy. Together with additional $K$-band data from \cite{Bloom991208} and
\cite{CharyHost}, we find a high extinction of $\AV=0.80\pm0.29$ for MW dust,
the weakly preferred model, and an excellent fit ($\chi^2$/dof=0.20). 

\emph{GRB 991216} After correcting for significant Galactic foreground
extinction, both \cite{Garnavich2000} and \cite{Halpern2000} find a SED well
approximated by a power-law, concurrent with no dust reddening, although these
authors do not discuss intrinsic reddening. With a larger database and more
colors, we find $\AV=0.13\pm0.08$ for the weakly preferred MW dust, still a
low value. This is consistent with the value $\AV=0.16\pm0.02$ that
\cite{Vreeswijk2005} find from a low-resolution spectrum. They also detect a
broad absorption feature centered at 2360 {\AA},  which can be interpreted as
a redshifted 2715 {\AA} bump, giving credence to our choice of MW dust as the
preferred model.

\emph{GRB 000131} For an SMC dust model, \cite{Andersen2000} find
$\AV=0.155\pm0.045$. We are not able to perform a free fit as the SED has only
three colors ($R_C$ and $V$ are affected by Lyman dampening), but fitting a
straight power-law shows the SED is consistent with no
extinction (i.e., no SED curvature is seen).

\emph{GRB 000301C} \cite{Jensen2001} find very low host extinction
($\AV=0.09\pm0.04$) for this burst using an SMC dust curve, and no solution
for MW and LMC dust.  \cite{RhoadsFruchter2001} derive a best fit result of
$\AV=0.09$ for SMC dust, find a slightly worse fit for LMC dust and a much
worse fit for MW dust. We also find negative extinction for MW dust, and also
low values for LMC and SMC dust ($\AV=0.12\pm0.06$ for SMC dust), completely
in concordance with the results of the other authors.

\emph{GRB 000911} Comparing their host galaxy SED with synthetic spectra of
extinguished galaxies, \cite{Masetti2005} find $\AV=0.32$ for SMC dust,
consistent with our result of $\AV=0.27\pm0.32$. The LMC result ($\AV=0.39$
compared to $\AV=0.27\pm0.30$) is also in agreement.

\emph{GRB 000926} For SMC dust, we find $\AV=0.15\pm0.07$, in good agreement
with \cite{Fynbo2001} who find $\AV=0.18\pm0.06$ for SMC dust. Several other
analyses \citep{Price2001, Sagar2001} disagree more with our
results, but \cite{Fynbo2001} employ a $UBVR_CI_CJHK$ spectral energy
distribution as we do, while the other studies have shorter baselines. Based
on metal column abundances from high-resolution spectra, \cite{Savaglio2003}
find a large $\AV=0.9$, which we can not confirm via the broad band 
photometry.

\emph{GRB 010222} Working with five-color SDSS photometry, \cite{Lee2001} find
low extinction ($\AV\leq0.057$ for $\beta=0.75$ fixed) for this burst with an
SMC dust curve, which is less than what we derive ($\AV=0.14\pm0.08$ for
$\beta=0.76$ and SMC dust). Our result is consistent with that of
\cite{Galama2003}, who find $\AV=0.1\pm0.02$.  \cite{Savaglio2003} find a
higher extinction of $\AV=0.7$ based on metal column abundances.

\emph{GRB 010921} Not finding a supernova bump in this nearby GRB,
\cite{Price2003} invoke $\AV\approx1$. While \cite{ZKH} do find a weak SN bump
and we derive the SED from this fit, we also find significant extinction of
$\AV=0.91\pm0.82$ for MW dust, the preferred model, thus validating the
analysis in \cite{Price2003}.

\emph{GRB 011121} \cite{Price2002} deduce a Galactic visual extinction of
$\AV=1.16\pm0.25$  along the line of sight by fitting the uncorrected data
points with a MW extinction law at $z=0$, which translates into  a Galactic
$\EBV=0.37\pm0.08$, assuming $\RV=$3.1. They could not constrain the host
extinction, however.  \cite{Garnavich2003} find a Galactic $\EBV=0.43\pm0.07$
and a spectral slope of $\beta=0.66\pm0.13$ between about 0.4 and 1.5 days
after the burst; a possible extinction in the host galaxy is not considered.
\cite{Greiner2003} adopt $\EBV=0.46$ based on the COBE data \citep{SFD}, find
no evidence for host extinction, and $\beta=0.62\pm0.05$ at 2.5 days after the
burst. They note, however, that the HI maps of \cite{DL1990} favor a lower
Galactic extinction of $\AV=0.9$ mag, i.e. $\EBV=0.29$ (assuming $\RV=$ 3.1).
We refitted the combined data set taken from \cite{Greiner2003}, adding
carefully selected (early and HST data) data from \cite{Garnavich2003} and
\cite{Price2002}.  In addition, we have now used $\EBV=0.29$ deduced via the
HI data \citep{DL1990}, in order to overcome the potential mixture of an
unknown source frame extinction and an uncertain Galactic extinction along the
line of sight. We omitted the $H$-band data  \citep{Greiner2003}, however,
since they worsen the fit. In doing so, we now derive a moderate host
extinction of $\AV\approx0.4$ mag (SMC dust). The MW dust model is not
preferred, although the low redshift of the GRB makes a differentiation among
the three dust models hard.

\emph{GRB 011211} Both \cite{Holland2002} and \cite{Jakobsson2003} find very
low extinction for this burst. \cite{Jakobsson2003} rule out MW dust and find
that SMC dust gives the best fit with $\AV=0.08\pm0.08$. Our result is that MW
dust is ruled out, LMC dust gives a low $\AV$ with a very bad fit, and SMC
dust gives a moderate $\AV=0.25\pm0.06$ but with $\beta=0\pm0.15$. 
Thus the results are comparable in principle.

\emph{GRB 020124} \cite{Hjorth2003} find $\AV\leq0.2$ by fixing $\beta>0.5$
for SMC dust, deeming $\beta<0.5$ unrealistic. Their SED is consistent with
ours, showing a strong curvature. They additionally employ a synthetic 
$B$-band point derived from extrapolating the spectrum and find
$\beta=0.31\pm0.43$ from a free fit of SMC dust, still higher than what we
derive (excluding the $B$ band, as it is affected by Lyman dampening),
although the results are identical within the large errors. \cite{Berger2002}
find $\AV=0$ to 0.9.

\emph{GRB 020405} Working with almost the same data set as we do, and
including X-ray observations, \cite{Masetti2003} find negligible (though
unspecified) dust extinction in this afterglow. On the other hand,
\cite{Stratta2005}, comparing the X-ray and the optical brightness, invoke a
gray extinction curve with a very high $\AV=2.3$. We find moderate extinction
$\AV\approx0.25$ for all three dust models.

\emph{GRB 020813} \cite{Savaglio2004} performed a very stringent analysis of
an early high-resolution spectrum of this afterglow and, via analysis of metal
column densities, derive $\AV\leq0.08$, $\AV\leq0.19$ and $\AV\leq0.18$ for
MW, LMC, and SMC dust, respectively. Our analysis gives a very good match,
with definite results: $\AV=0.01\pm0.08$, $\AV=0.19\pm0.12$ and
$\AV=0.12\pm0.07$, respectively.  Furthermore, we also come to the conclusion
that SMC dust is the preferred model. To make their unextincted spectral slope
more compatible with standard models, \cite{Li2003} invoke $\EBV=0.05$ (which
converts to $\AV=0.15$). \cite{Covino2003} deduce $\AV=0.12\pm0.04$, but for
the Galactic extinction curve of \cite{CCM}.

\emph{GRB 021004} For an SMC dust curve, \cite{Holland2003} derive
$\AV=0.26\pm0.04$, coupled with $\beta=0.39\pm0.12$. This work finds a lower
$\AV=0.14\pm0.05$ coupled with a higher $\beta=0.67\pm0.14$. This is for the
free fit, a wind model with the cooling break blueward of the optical gives
$\beta=0.38$ fixed and $\AV=0.20\pm0.02$, which is comparable. Fitting an SED
derived on October 11.67, \cite{Fynbo2005} find exactly this result
($\AV=0.20\pm0.02$) for SMC dust, associated with $\beta=0.42 \pm0.06$. They
find higher extinction for LMC dust and a worse fit, and negative extinction
for MW dust, just as we do.

\emph{GRB 021211} \cite{Fox2003} derive extinction values for this GRB by
fixing $\beta$. As our light curve fitting finds a very blue afterglow with
$\beta_0=0.10\pm0.09$ (i.e., almost flat), these values can not be compared
adequately. \cite{Fox2003} find up to $\AV=0.64$, while we only derive small
negative extinction for all dust models. Since these authors have derived
their spectral slope $\beta\approx1$ from an early $B-K_S$ color, this may
imply color evolution \citep[cf.][]{Nysewander05}. \cite{Holland2004} derive
an unextincted slope of $\beta_0=0.69\pm0.14$ at 0.87 days after the burst in
the observer frame and an upper limit on the source frame extinction of
$\AV\leq0.08$.

\emph{GRB 030226} \cite{Klose2004} find negligible extinction for this
afterglow based on their multicolor data set, 
in contrast to \cite{Pandey2004} who, on theoretical grounds, argue
for a fixed $\beta=0.55$ and thus derive small extinction from their
$\beta_0=1$. We utilize both data sets, and find a result concurrent with
\cite{Klose2004}, MW and LMC dust are ruled out, and the reddening using SMC
dust ($\AV=0.06\pm0.06$) is negligible within errors.

\emph{GRB 030227} \cite{CastroTirado2003} were not able to conclusively fit
their SED. Their $\beta_0=1.25\pm0.14$ is in perfect agreement with the result
we find, $\beta_0=1.24\pm0.13$. While the data set is sparse, it is, with the
exception of one data point, from \cite{CastroTirado2003} exclusively. The
steep $\beta_0$ supports dust reddening, and the unextinguished $\beta=0.78$,
a reasonable value, further supports our findings of
$\AV=0.38\pm1.81$, which could explain the faintness of this not too distant
\citep[$z=1.39$, ][]{Watson2003} afterglow. The sparse data points of the SED
and their large errors lead to very large errors in $\beta$ and $\AV$, however.

\emph{GRB 030323} The SED of this burst is unusual. The $B$ band can not be
included due to Lyman dampening, and the spectrum (a strong DLA) shows that
the $V$ band is also unreliable. With the exception of the $K$ band, the SED
shows a strong curvature without being very steep. As there is no reason to
remove the $K$ band point (based on three photometric points, and not two each
as in $J$ and $H$), we tried  to fit the SED with it (removing it results in
very good fits that have $\beta\leq0$). As the data points have small errors,
the result is a very bad fit ($\chi^2_{\nu}\approx6$ for two degrees of
freedom). \cite{Vreeswijk2004} do not fit the light curves, but analyse the
SED at different post-burst times from observations at near-identical
epochs. They fix $\beta=0.28$,  derived from the $\alpha$-$\beta$
relations. Using this value, they find $\AV=0.5$, $\AV=0.25$ and $\AV=0.16$
for MW, LMC and SMC dust curves, respectively. Fixing $\beta$ to this value
and fitting our SED yields $\AV=0.70$, $\AV=0.41$ and $\AV=0.26$ for MW, LMC
and SMC dust curves, respectively. While these values are higher than those in
\cite{Vreeswijk2004}, they are comparable if conservative errors are assumed.
However, the $K$-band data is a strong outlier in these fits.

\emph{GRB 030328} After Paper II went into press, \cite{Maiorano2006}
presented a large amount of $UBVR_CI_C$ data on this burst that was not
available to us before. We refitted the optical light curves and derive
the following parameters which supersede those presented in Table 1 of
Paper II: $\chi^2/\tn{dof}=0.72$ (1.34 in Paper II), dof = 81 (18 in
Paper II), $m_c=19.54\pm0.31$ mag ($20.61\pm0.23$ in Paper II), $\alpha_1
=0.61\pm0.13$ ($0.87\pm0.04$ in Paper II), $\alpha_1=1.41\pm0.12$
($1.54\pm0.11$ in Paper II), $t_b=0.29\pm0.06$ days after the burst
($0.60\pm0.10$ in Paper II), $n=3.19\pm2.74$ (fixed to 10 in Paper II).
The host galaxy magnitude remains unchanged. (See Paper II for the
definitions of these values.) These changes do not influence the
conclusions of Paper II. We note that the value for the break smoothness
parameter $n$ is fully in agreement with the possible correlation
between $\alpha_1$ and $n$ (Figure 8 of Paper II), and is almost
identical to the value pair derived for GRB 010222.

From the Fe II column density derived from the optical spectrum,
\cite{Maiorano2006} estimate $\AV\leq 0.1$ mag, in agreement with our
result $\AV=0.05\pm0.15$ mag.

\emph{GRB 030329} The SED of this burst was derived in an alternate way
compared to all other SEDs.  The rebrightening episodes starting at one day
are found to be achromatic over several days. The high data quality makes it
possible to shift the light curves to a common magnitude, the amount of shift
is used to construct the SED, assuming conservative errors. This method is
independent of the light curve fit, which is very complicated in the case of
GRB 030329 (cf. Paper II).  \cite{Bloom2004} find $\AV=0.94\pm0.24$, coupled
with $\beta=0.11\pm0.22$, for MW dust. Our result is: $\AV=0.54\pm0.22$,
coupled with $\beta=0.30\pm0.22$, which is comparable within errors.  For
$\beta=0.5$ fixed, \cite{Bloom2004} derive $\AV\leq0.3$, comparable to
$\AV=0.34\pm0.04$ which we derive when fixing $\beta$ at this value.

\emph{GRB 030429} \cite{Jakobsson2004} rule out MW dust (as do we) and find
$\AV=0.30\pm0.06$ for SMC dust at 0.548 days after the burst, slightly lower
than our value of $\AV=0.40\pm0.10$. They also find higher extinction for LMC
dust, but the fit is much worse, contrary in part to our result. We find much
lower source frame extinction for LMC dust, but here also the quality of the
fit is bad.

\emph{XRF 030723} \cite{FynboXRF030723} give $\AV\leq0.5$ for SMC dust and
$z=0.3$, completely concurrent with our result of $\AV=0.28\pm0.24$ for SMC
dust and under the assumption of a redshift of $z=0.35$.

\emph{GRB 040924} \cite{SilveyGCN} use their own data and data from the GCN
archives to derive $\beta_0=0.61\pm0.08$, slightly lower than the value we
find ($\beta_0=0.80\pm0.03$). They do not fit the SED with a dust model.
\cite{Soderberg2006} derive $\beta_0\approx 0.7$ and argue that for the
standard blastwave model, $\beta\geq 0.5$ and thus $\AV\leq0.16$ mag. This
is in agreement with our result $\AV=0.16\pm0.44$ mag.

\emph{GRB 041006} \cite{Soderberg2006} find $\beta_0\approx 0.5$ and, to
keep $p\geq 2$, assume negligible host galaxy extinction. This is in
agreement with our result $\AV=0.11\pm0.23$, but we note that the shape
of the SED is indicative of a slight amount of dust. The derived
$\beta=0.36\pm0.27$ and the shallow post-break light curve decay slope
$\alpha_2=1.30\pm0.02$ (Paper II) are both indicative of a hard electron
spectrum with $p\leq 2$.


\begin{deluxetable}{l r c | c c c | c c c | c c c}
\rotate
\tablecolumns{12}
\tabletypesize{\scriptsize}
\tablewidth{0pc}
\tablecaption{Results of the SED fitting}
\tablehead{
\colhead{GRB}   &
\colhead{\#\tablenotemark{a}} &
\colhead{Filters\tablenotemark{b}} &
\multicolumn{3}{c}{MW Dust} &
\multicolumn{3}{c}{LMC Dust} &
\multicolumn{3}{c}{SMC Dust}\\
\multicolumn{3}{c}{} &
\colhead{$\chi^2_{dof}$}  &
\colhead{$\beta$}  &
\colhead{$\AV$} &
\colhead{$\chi^2_{dof}$}  &
\colhead{$\beta$}  &
\colhead{$\AV$} &
\colhead{$\chi^2_{dof}$}  &
\colhead{$\beta$}  &
\colhead{$\AV$}}
\startdata
970508  &               &$BVR_CI_CK_S$                  &       2.73    & $0.32\pm0.15$ & $0.38\pm0.11$ &       2.00    & $0.11\pm0.20$ & $0.55\pm0.15$ &       1.14    & $0.00\pm0.22$ & $0.61\pm0.15$ \\
971214  &               &$R_CI_CJK$                                                                     &       0.32    & $2.26\pm0.81$ & $-1.44\pm0.72$&       0.14    & $-1.26\pm0.95$& $1.06\pm0.52$ &       0.09    & $-0.50\pm0.58$& $0.44\pm0.21$ \\
980519\tablenotemark{c} &       1       &$UBVR_CI_C$&   0.61    & $1.11\pm0.11$ & $-0.03\pm0.05$&       0.73    & $1.10\pm0.26$ & $-0.02\pm0.12$&       0.01    & $0.44\pm0.54$ & $0.22\pm0.19$ \\
980703  &               &$BVR_CI_CJHK$                                                  &       2.28    & $0.57\pm0.92$ & $1.93\pm0.91$ &       2.10    & $0.57\pm0.85$ & $1.85\pm0.81$ &       2.33    & $1.05\pm0.66$ & $1.32\pm0.59$ \\
990123  &       2       &$UBVR_CI_CHK$                                                  &       0.21    & $0.62\pm0.09$ & $-0.06\pm0.07$&       0.35    & $0.50\pm0.20$ & $0.04\pm0.13$ &       0.20    & $0.46\pm0.12$ & $0.04\pm0.05$ \\
990510  &       3       &$BVR_CI_C$                                                                     &       0.03    & $0.89\pm0.11$ & $-0.05\pm0.04$&       0.01    & $1.12\pm0.31$ & $-0.15\pm0.15$&       0.58    & $0.30\pm0.69$ & $0.18\pm0.24$ \\
991208  &               &$BVR_CI_CK$                                                            &       0.20    & $0.23\pm0.37$ & $0.80\pm0.29$ &       0.17    & $0.07\pm0.43$ & $0.93\pm0.34$ &       0.26    & $0.19\pm0.40$ & $0.76\pm0.28$ \\
991216  &       4       &$BVR_CI_CJHK$                                                  &       0.12    & $0.38\pm0.11$ & $0.13\pm0.08$ &       0.16    & $0.32\pm0.15$ & $0.18\pm0.11$ &       0.29    & $0.30\pm0.17$ & $0.18\pm0.13$ \\
000131\tablenotemark{d} & &$I_CHK_S$            &       $<$0.01&$0.66\pm0.34$ & 0                                               &$<$0.01& $0.66\pm0.34$ & 0                                             &$<$0.01& $0.66\pm0.34$ & 0                                     \\
000301C &       5       &$BVR_CI_CJK$                                                           &       2.48    & $0.88\pm0.07$ & $-0.03\pm0.05$&       2.03    & $0.59\pm0.19$ & $0.16\pm0.11$ &       1.14    & $0.59\pm0.12$ & $0.12\pm0.06$ \\
000911  &               &$BVR_CI_CJK_S$                 &       0.26    & $0.75\pm0.26$ & $0.20\pm0.22$ &       0.28    & $0.67\pm0.36$ & $0.27\pm0.30$ &       0.33    & $0.65\pm0.40$ & $0.27\pm0.32$ \\
000926  &       6       &$BVR_CI_CJHK$                                                  &       1.02    & $1.43\pm0.07$ & $\-0.07\pm0.05$&1.67  & $1.29\pm0.20$ & $0.04\pm0.12$ &       0.37    & $1.01\pm0.16$ & $0.15\pm0.07$ \\
010222  &       7       &$UBVR_CI_CJK$                                                  &       1.35    & $1.18\pm0.07$ & $-0.03\pm0.04$&       1.36    & $1.02\pm0.20$ & $0.07\pm0.10$ &       0.58    & $0.76\pm0.22$ & $0.14\pm0.08$ \\
010921  &               &$UBVR_Cr^*i^*$                                                                         &       0.07    & $0.81\pm1.21$ & $0.91\pm0.82$ &       0.06    & $0.03\pm1.88$ & $1.44\pm1.29$ &       0.03    & $-1.01\pm2.71$& $1.91\pm1.66$ \\
011121  &       8       &$UBVR_CI_CJK$                                                  &       0.55    & $0.55\pm0.15$ & $0.49\pm0.18$&        0.54    & $0.55\pm0.15$ & $0.47\pm0.17$&        0.49    & $0.61\pm0.13$ & $0.39\pm0.14$\\
011211  &       9       &$BVR_CI_CJK$                                                           &       2.03    & $0.88\pm0.09$ & $-0.24\pm0.06$&       7.12    & $0.38\pm0.24$ & $0.11\pm0.14$ &       1.89    & $0.00\pm0.15$ & $0.25\pm0.06$ \\
020124  &       10&$R_CI_CJK_S$                         &       0.14    & $1.29\pm0.65$ & $-0.44\pm0.59$&       0.15    & $-0.49\pm1.77$& $1.34\pm0.98$ &       0.01    & $0.11\pm0.85$ & $0.28\pm0.33$ \\
020405  &       11&$BVR_CI_CJHK_S$              &       1.82    & $0.96\pm0.20$ & $0.14\pm0.18$ &       1.75    & $0.91\pm0.22$ & $0.19\pm0.20$ &       1.75    & $0.94\pm0.19$ & $0.15\pm0.16$ \\
020813  &       12&$UBVR_CI_CJHK$                                                       &       1.62    & $1.03\pm0.11$ & $0.01\pm0.08$ &       1.10    & $0.74\pm0.20$ & $0.19\pm0.12$ &       0.95    & $0.81\pm0.14$ & $0.12\pm0.07$ \\
021004  &               &$BVR_CI_CJHK_S$                &       1.62    & $1.31\pm0.10$ & $-0.20\pm0.08$&       1.57    & $0.56\pm0.22$ & $0.27\pm0.11$ &       0.69    & $0.67\pm0.14$ & $0.14\pm0.05$ \\
021211  &         &$BVR_CI_CJHK_S$              &       1.16    & $0.15\pm0.23$ & $-0.04\pm0.15$&       1.13    & $0.22\pm0.32$ & $-0.09\pm0.23$&       1.05    & $0.38\pm0.40$ & $-0.19\pm0.27$\\
030226  &       13&$BVR_CI_CJHK$                                                        &       0.15    & $0.77\pm0.06$ & $-0.06\pm0.04$&       0.57    & $0.78\pm0.18$ & $-0.05\pm0.11$&       0.32    & $0.57\pm0.12$ & $0.06\pm0.06$ \\
030227  &               &$BR_CHK$                                                                                                               &       1.41    & $0.78\pm2.17$ & $0.38\pm1.81$ &       0.74    & $2.23\pm1.15$ & $-0.76\pm0.87$&       0.48    & $1.89\pm0.67$ & $-0.44\pm0.43$\\
030323  &               &$R_CI_CJHK$                                                            &       5.91    & $1.43\pm0.35$ & $-0.32\pm0.31$&       6.35    & $1.30\pm0.54$ & $-0.12\pm0.28$&       6.41    & $1.14\pm0.31$ & $-0.02\pm0.11$\\
030328  &       14&$UBVR_CI_C$                                                                   &       0.23    & $0.51\pm0.06$ & $0.00\pm0.03$&       0.22    & $0.49\pm0.15$ & $0.01\pm0.07$& 0.18 & $0.36\pm0.45$& $0.05\pm0.15$ \\
030329  &               &$UBVR_CI_CJH$                                                  &       0.10    & $0.30\pm0.22$ & $0.54\pm0.22$ &       0.09    & $0.32\pm0.21$ & $0.50\pm0.20$ &       0.06    & $0.41\pm0.17$ & $0.39\pm0.15$ \\
030429  &       15&$VR_CI_CJHK_S$                       &       2.60    & $1.51\pm0.09$ & $-0.28\pm0.07$&       7.72    & $1.11\pm0.33$ & $0.05\pm0.19$ &       1.79    & $0.22\pm0.24$ & $0.40\pm0.10$ \\
030723\tablenotemark{c} &         &$UBVR_CiJ_SHK_S$     &       0.16    & $0.58\pm0.26$ & $0.48\pm0.29$ &       0.15    &       $0.58\pm0.25$   &       $0.42\pm0.28$   &       0.19    & $0.66\pm0.21$ & $0.32\pm0.22$ \\
040924  &               &$VR_CI_CK$                                                                     &       0.10    & $0.59\pm0.61$ & $0.21\pm0.62$ &       0.10    & $0.58\pm0.64$ & $0.22\pm0.62$ &       0.09    & $0.63\pm0.48$ & $0.16\pm0.44$ \\
041006  &       16&$BVR_CI_CH$                                                          &       0.06    & $0.36\pm0.27$ & $0.11\pm0.23$ &       0.05    & $0.32\pm0.33$ & $0.14\pm0.28$ &       0.05    & $0.34\pm0.30$ & $0.12\pm0.23$ \\
\enddata
\tablenotetext{a}{denotes the corresponding number in the sample
constructed in Paper II.}
\tablenotetext{b}{Filters that are not used for the fit (e.g,
due to Lyman $\alpha$ dampening) are not included. The degrees of
freedom of the fit are always number of filters minus three,
except for GRB 000131.}
\tablenotetext{c}{For GRB 980519 and XRF 030723, redshift
estimates were employed (see \kref{data}).}
\tablenotetext{d}{With only three viable colors, GRB 000131
could not be fit with three free parameters. The fit was
performed without extinction. No dust preference can be derived.
The fit has one degree of freedom.}
\label{tabALL}
\end{deluxetable}

\newpage\clearpage

\begin{deluxetable}{l c c c c}
\tablecolumns{5}
\tablewidth{0pc}
\tablecaption{The SED Golden Sample constructed from Table \ref{tabALL}}
\tablehead{
\colhead{GRB}   &
\colhead{Dust\tablenotemark{a}} &
\colhead{$\chi^2_{dof}$}  &
\colhead{$\beta$}  &
\colhead{$\AV$}}
\startdata
970508  &       \bf{\emph{M}}   &       2.73    & $0.32\pm0.15$ & $0.38\pm0.11$ \\
990123  &       \bf{\emph{S}}   &       0.20    & $0.46\pm0.12$ & $0.04\pm0.05$ \\
991208  &       M                               &       0.20    & $0.23\pm0.37$ & $0.80\pm0.29$ \\
991216  &       M                               &       0.12    & $0.38\pm0.11$ & $0.13\pm0.08$ \\
000131  & $-$                   &$<$0.01& $0.66\pm0.34$ & 0                                     \\
000301C &       \bf{\emph{S}}   &       1.14    & $0.59\pm0.12$ & $0.12\pm0.06$ \\
000911  &       M                               &       0.26    & $0.75\pm0.26$ & $0.20\pm0.22$ \\
000926  &       \bf{\emph{S}}   &       0.37    & $1.01\pm0.16$ & $0.15\pm0.07$ \\
010222  &       \bf{\emph{S}}   &       0.58    & $0.76\pm0.22$ & $0.14\pm0.08$ \\
011121  &       S                               &       0.49    & $0.61\pm0.13$ & $0.39\pm0.14$\\
020405  &       S                               &       1.75    & $0.94\pm0.19$ & $0.15\pm0.16$ \\
020813  &       S                               &       0.95    & $0.81\pm0.14$ & $0.12\pm0.07$ \\
021004  &       \bf{\emph{S}}   &       0.69    & $0.67\pm0.14$ & $0.14\pm0.05$ \\
030226  &       \bf{\emph{S}}   &       0.32    & $0.57\pm0.12$ & $0.06\pm0.06$ \\
030328  &       S               &       0.18    & $0.36\pm0.45$ & $0.05\pm0.15$ \\
030329  &       S                               &       0.06    & $0.41\pm0.17$ & $0.39\pm0.15$ \\
030429  &       \bf{\emph{S}}   &       1.79    & $0.22\pm0.24$ & $0.40\pm0.10$ \\
040924  &       S                               &       0.09    & $0.63\pm0.48$ & $0.16\pm0.44$ \\
041006  &       M                               &       0.06    & $0.36\pm0.27$ & $0.11\pm0.23$ \\
\enddata
\label{tabGold}
\tablenotetext{a}{This represents which dust model is used in
the fit. Bold \textbf{\emph{M}} and \textbf{\emph{S}}
mean a MW and SMC dust fit, respectively, with strong
arguments favoring this fit. M and S mean a MW and SMC dust fit,
respectively, with weak arguments favoring this fit. If no
letter is given, then $\AV=0$, and no conclusions can be drawn
about the dust model.}
\end{deluxetable}

\newpage\clearpage

\begin{table}[!t]
\scriptsize
\caption{$\NH$ along the line-of-sight to the GRBs}
\label{Xraydata}
\begin{tabular}{|l|l|c|c|}\hline
GRB    & $\NH$ ($10^{21}\tn{cm}^{-2}$) & Method\tablenotemark{a} & Reference \\ \hline\hline
970508 &   6$^{+10}_{-5}$    & X & \cite{GalamaWijers2001} \\
990123 & 5.4$^{+9.4}_{-2.7}$ & X & \cite{GalamaWijers2001} \\
991216 & 6.8$^{+6.3}_{-4.4}$ & X & \cite{Ballantyne2002} \\
000301C& 1.6$^{+3.4}_{-1.1}$ & O & \cite{Jensen2001} \\
000926 &   4$^{+3.5}_{-2.5}$ & X & \cite{Piro2001} \\
010222 &  12$^{+7}_{-6}$     & X & \cite{Stratta2004} \\
020405 & 4.7$^{+3.7}_{-3.7}$ & X & \cite{Mirabal2003} \\
021004 &$<0.1$               & X & \cite{Moller2002} \\
030226 & 3.2$^{+1.5}_{-1.5}$ & X & \cite{Klose2004} \\
030329\tablenotemark{b} &$<0.2$ & X & \cite{Tiengo2004} \\ 
030429 & 4.0$^{+2.3}_{-1.5}$ & O & \cite{Jakobsson2004} \\
041006 & 3.2$^{+0.16}_{-0.16}$ & X & \cite{Butler2005} \\ \hline
\end{tabular}
\tablenotetext{a}{Method: X -- via X-ray data; O -- via Lyman $\alpha$ (optical)}
\tablenotetext{b}{The value for GRB
030329 has been transformed into an upper limit.}
\end{table}

\newpage\clearpage

\begin{table}[!t]
\caption{Redshift corrections of the afterglows listed in Table~\ref{tabGold}}
\begin{tabular}{|l r| l r| l r| l r|}\hline
 & & & & & & & \\[-2mm]
GRB & $dR_C$\tablenotemark{a}& GRB & $dR_C$ & GRB & $dR_C$ & GRB & $dR_C$ \\[1mm] \hline
970508 &  $-$0.16 &  000301C & $-$2.09 & 020405 &        0.77 &  030329 &    3.87 \\
990123 &  $-$1.22 &  000911 &  $-$0.49 & 020813 & $-$0.83 &  030429 & $-$3.56 \\
991208 &  $-$0.35 &  000926 &  $-$2.36 & 021004 & $-$2.55 &  040924 &    0.12 \\ 
991216 &  $-$0.27 &  010222 &  $-$1.32 & 030226 & $-$1.84 &  041006 &    0.64 \\
000131 &  $-$3.64 &  011121 &     2.08 & 030328 & $-$1.09 &         &         \\ \hline
\end{tabular}
\label{tabzCorr}
\tablenotetext{a}{$dR_C$ is the second term of Equation \ref{delta1}.}
\end{table}

\newpage\clearpage

\begin{table}[!t]
\caption{Absolute magnitudes $\MB$ of the afterglows}
\begin{tabular}{|l|c|c|c|}\hline
&&&\\[-2mm]
GRB     & Group\tablenotemark{a}& $\MB$ (at one day\tablenotemark{b})  
        &       $\MB$ (at four days\tablenotemark{b})
\\[1mm] \hline
970508  &\emph{II}& $-21.85^{+0.14}_{-0.15}$  & $-22.53^{+0.14}_{-0.15}$  \\
990123  &\emph{I} & $-23.21^{+0.03}_{-0.05}$  & $-20.92^{+0.03}_{-0.05}$  \\
991208  &         & $\cdots$                  & $-23.25^{+0.39}_{-0.41}$  \\
991216  &\emph{I} & $-24.89^{+0.11}_{-0.10}$  & $-22.72^{+0.11}_{-0.10}$  \\
000131  &         & $\cdots$                                                            & $\cdots$                  \\
000301C &         & $-25.01^{+0.09}_{-0.09}$    & $-23.35^{+0.09}_{-0.09}$  \\
000911  &         & $\cdots$                                                            & $-21.13^{+0.26}_{-0.30}$  \\
000926  &\emph{I} & $-25.12^{+0.08}_{-0.09}$  & $-21.80^{+0.08}_{-0.09}$  \\
010222  &\emph{I} & $-23.73^{+0.08}_{-0.07}$  & $-21.85^{+0.08}_{-0.07}$  \\
011121  &\emph{II}& $-22.70^{+0.19}_{-0.18}$  & $-18.92^{+0.19}_{-0.18}$  \\
020405  &         & $-22.39^{+0.20}_{-0.23}$  & $-20.03^{+0.20}_{-0.23}$  \\
020813  &         & $-23.23^{+0.08}_{-0.09}$  & $-21.13^{+0.08}_{-0.09}$  \\
021004  &         & $-25.58^{+0.04}_{-0.04}$  & $-24.07^{+0.04}_{-0.04}$  \\
030226  &\emph{II}& $-22.99^{+0.08}_{-0.09}$  & $-19.54^{+0.08}_{-0.09}$  \\
030328  &         & $-21.95^{+0.12}_{-0.04}$  & $\cdots$                  \\
030329  &\emph{II}& $-23.93^{+0.19}_{-0.21}$  & $-22.20^{+0.19}_{-0.21}$  \\
030429  &         & $-24.92^{+0.15}_{-0.16}$  & $-20.75^{+0.15}_{-0.16}$  \\
040924  &         & $-19.84^{+0.17}_{-0.65}$  & $\cdots$                  \\
041006  &         & $-20.76^{+0.13}_{-0.30}$  & $-18.89^{+0.13}_{-0.30}$  \\ \hline
\end{tabular}
\tablenotetext{a}{Denotes membership in \emph{group I} or \emph{group II}
of \cite{GB2005}}
\tablenotetext{b}{after the GRB, assuming at $z$=1}
\label{tabMB}
\end{table}

\newpage\clearpage

\begin{figure*}[t]
\epsfig{file=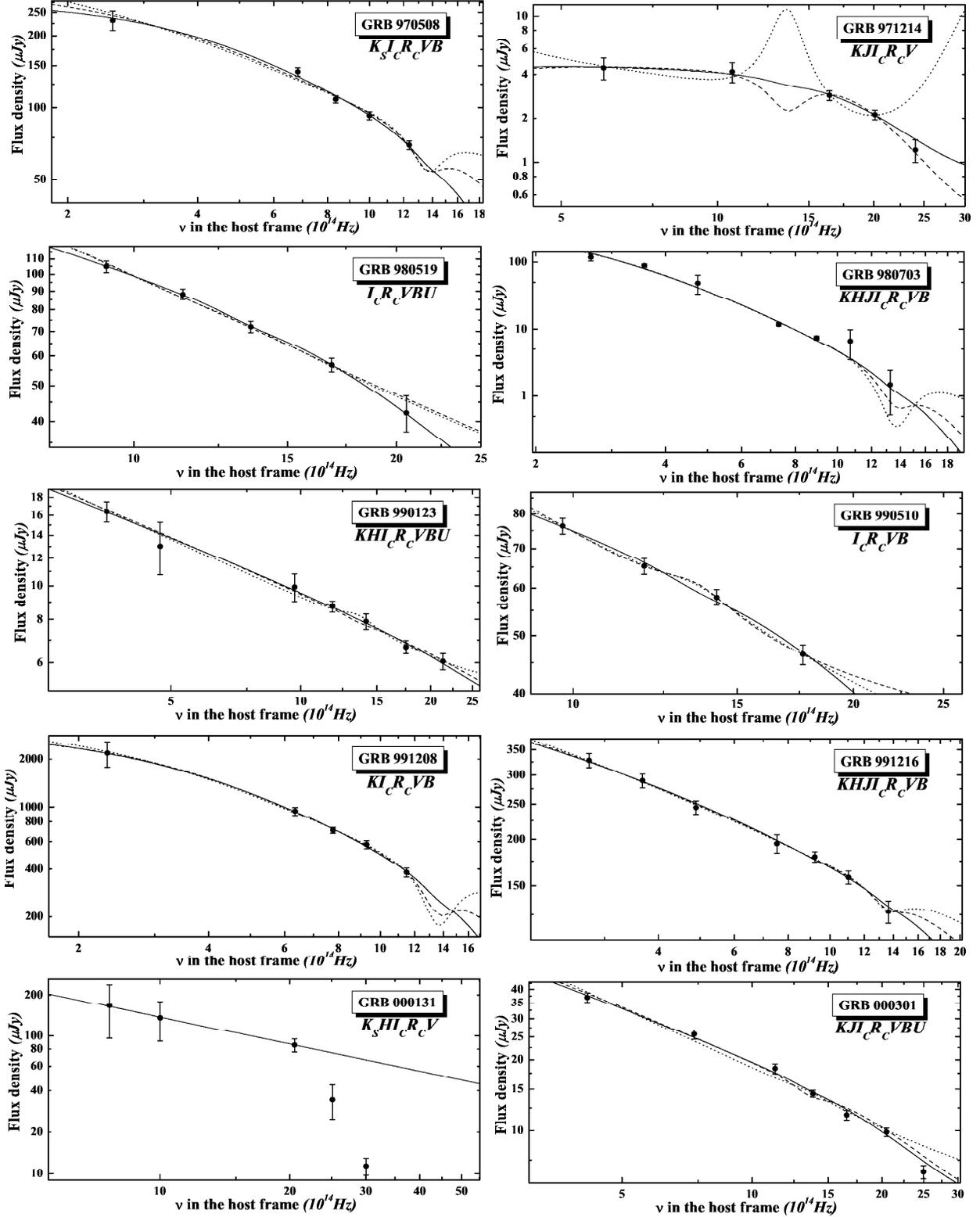,width=1\textwidth}
\caption[]{The spectral energy distributions of all GRB afterglows presented in
Table \ref{tabALL}. Included are the fits to a Milky Way (MW) dust model (dotted),
a Large Magellanic Cloud (LMC) dust model (dashed) and a Small Magellanic Cloud
dust model (solid). This includes fits with unphysical results (e.g., negative
extinction). Beneath the designation of the GRB, the filters are given that the
SED is based on. The absolute value of the flux density is not significant,
depending upon the fit and the break time (cf. \kref{data}).}
\label{SEDs}
\end{figure*}

\newpage\clearpage

\begin{figure*}[t]
\epsfig{file=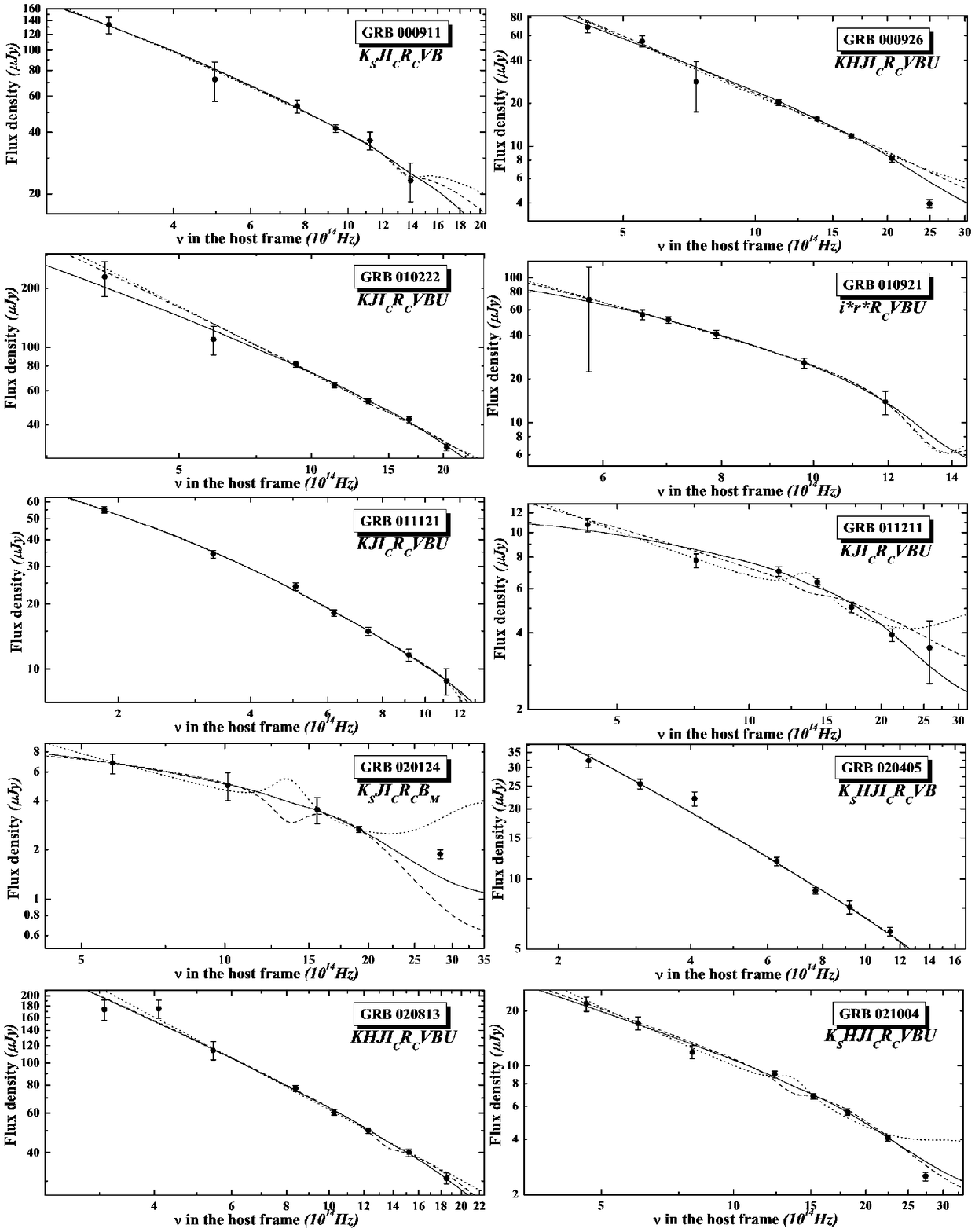,width=1\textwidth}
\addtocounter{figure}{-1}
\caption[]{continued.}
\end{figure*}

\newpage\clearpage

\begin{figure*}[t]
\epsfig{file=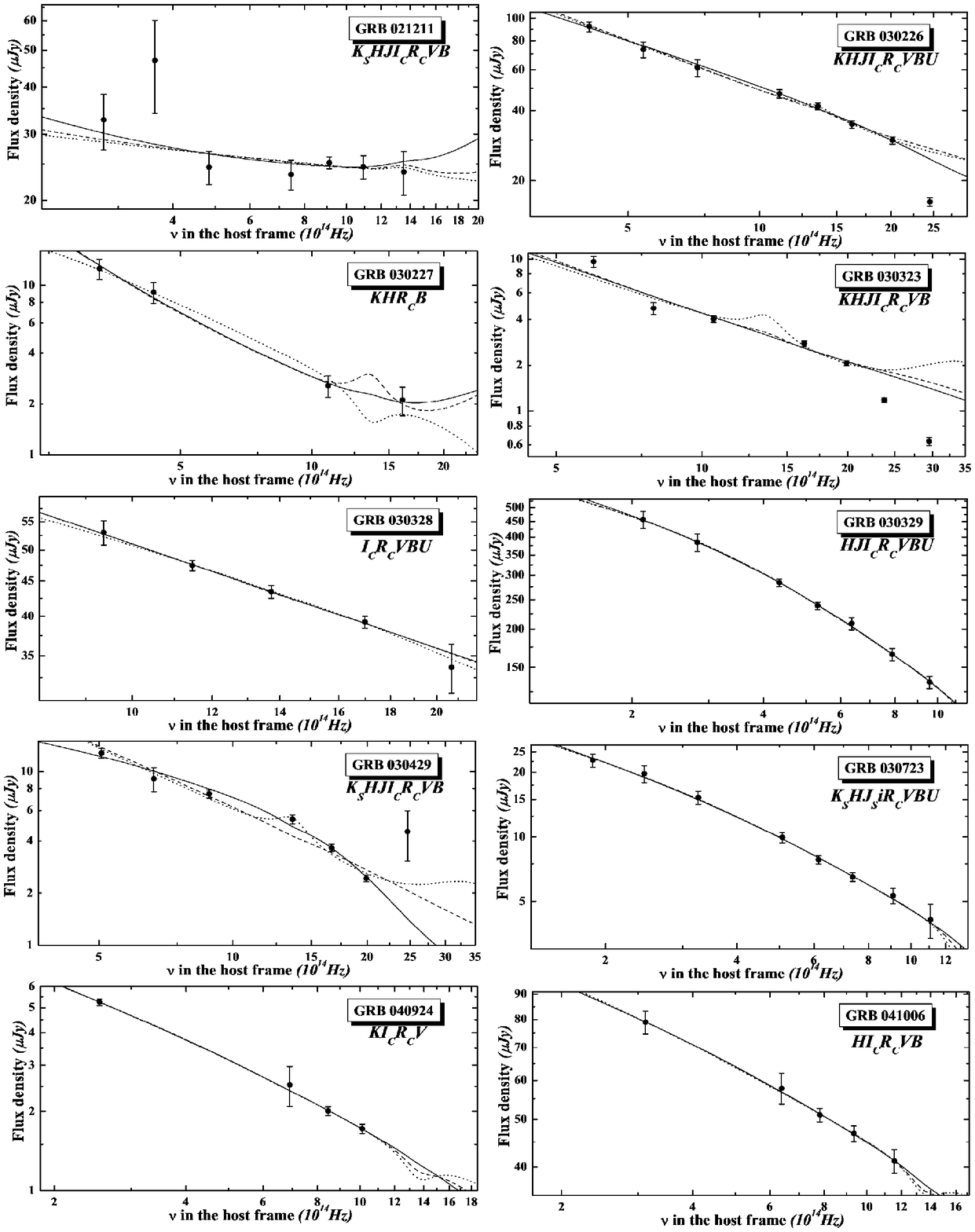,width=1\textwidth}
\addtocounter{figure}{-1}
\caption[]{continued.}
\end{figure*}

\newpage\clearpage

\begin{figure}[!t]
\epsfig{file=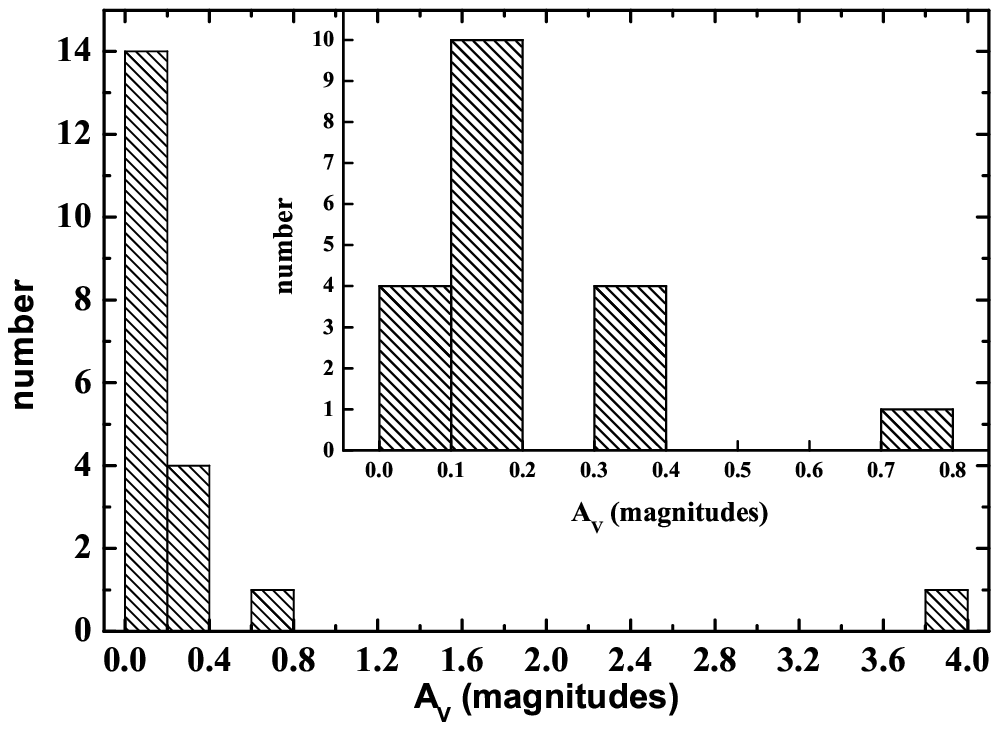,width=1\textwidth}
\caption[]{The distribution of the derived host galaxy visual extinction $\AV$ in the
source frame for the bursts of the Golden Sample (Table \ref{tabGold}). The
data point at $\AV=3.8$ is the lower limit on the visual extinction derived for GRB
970828 \citep[]{Djorgovski2001}. There are no afterglows with $\AV\gtrsim0.8$, creating
a Dark Burst Desert. The inset shows a zoom into the $\AV\lesssim0.8$ region with smaller
bins. The distribution peaks around $\AV=0.15$ mag, showing that there is typically a
small but definite amount of extinction along the line of sight to GRB afterglows.}
\label{A_VHist}
\end{figure}

\newpage\clearpage

\begin{figure}[ht]
\epsfig{file=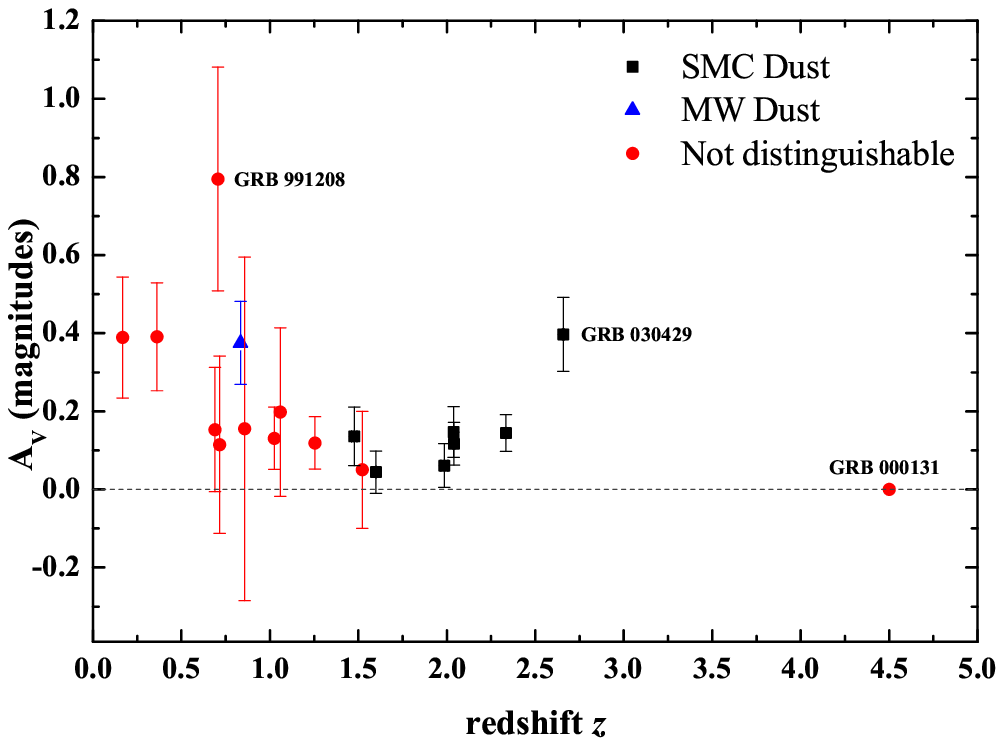,width=1\textwidth}
\caption[]{The derived host galaxy visual extinction $\AV$ in the source frame for
the Golden Sample bursts (Table \ref{tabGold}) plotted as a function of the redshift $z$
\citep[taken from the compilation of][]{FB2005}. A clear trend toward lower extinctions
at higher redshifts is visible. The preferred dust models are indicated by symbol form
(and color). Definite MW dust is only found for GRB 970508. For low redshifts, the
distinction between the three dust models disappears. Extinction was fixed to $\AV=0$
for GRB 000131.}
\label{A_Vtoz}
\end{figure}

\newpage\clearpage

\begin{figure}[!t]
\epsfig{file=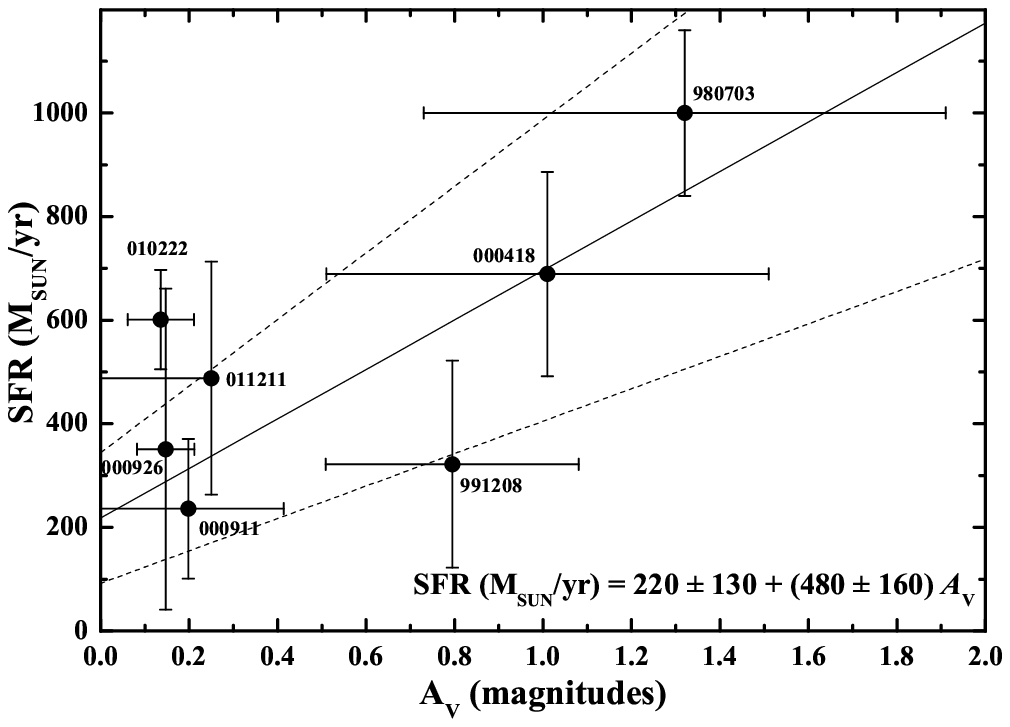,width=1\textwidth}
\caption[]{The star formation rate (SFR) derived from the submm flux density at 850
$\mu m$ for selected GRB host galaxies as a function
of the visual extinction $\AV$ along the line of sight. There is a correlation, although
the very large error bars prevent a deeper analysis. GRB 010222 lies beyond the
1$\sigma$ error region of the fit and was therefore not included in the fit.}
\label{submm}
\end{figure}

\newpage\clearpage

\begin{figure}[!t]
\epsfig{file=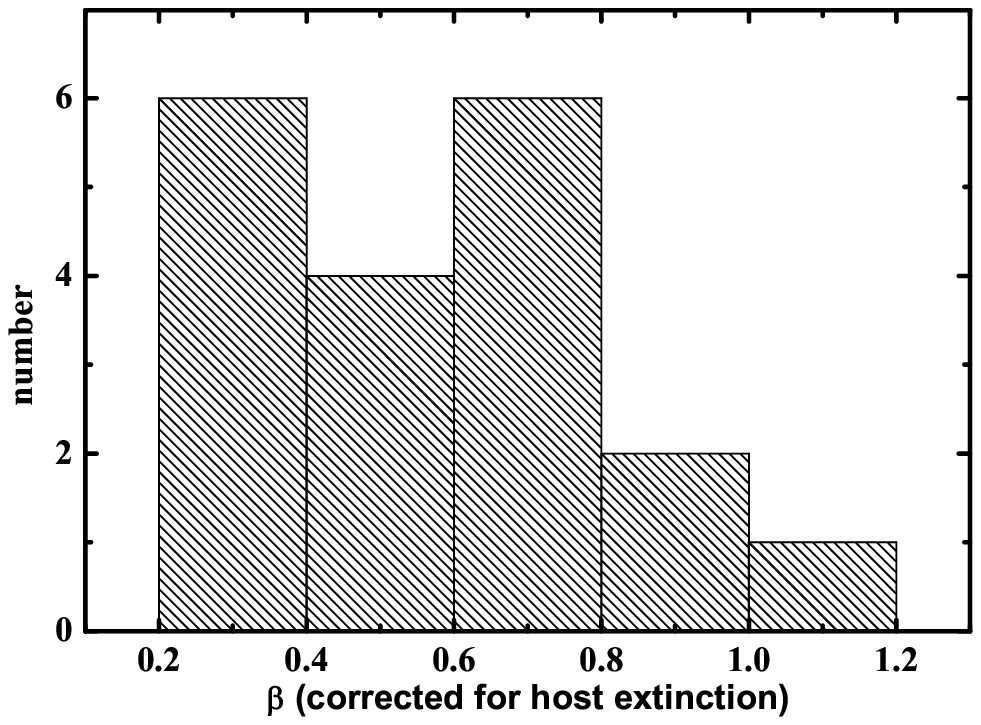,width=1\textwidth}
\caption[]{The distribution of the derived intrinsic spectral slopes $\beta$ for the
Golden Sample bursts (Table \ref{tabGold}). A broad peak around $\beta=0.7$ is visible
(implying $p=2.4$ for a wind environment with the cooling break lying blueward of the
optical), but several GRBs (GRB 970508, GRB 991208, GRB 991216, GRB 030328, GRB 030429,
GRB 041006) have $\beta\lesssim0.4$ (implying $p\lesssim1.8$ for the case stated
above).}
\label{betaHist}
\end{figure}

\newpage\clearpage

\newpage\clearpage

\begin{figure}[!t]
\epsfig{file=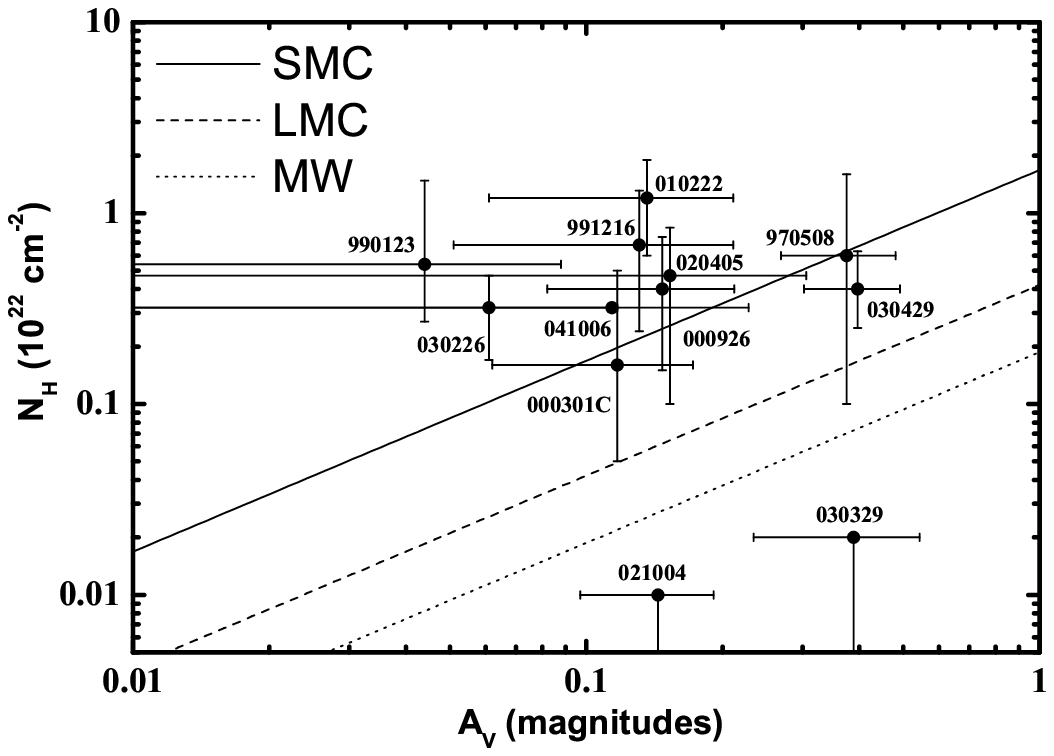,width=1\textwidth}
\caption[]{Dust-to-gas ratios in the host galaxies along the line of sight to the GRB
of the Golden Sample (Table \ref{tabGold}), with $\NH$ values obtained from the
literature (Table \ref{Xraydata}) and the $\NH$-$\AV$ relations taken from \cite{Bohlin1978},
\cite{PS1995} and \cite{Hopkins2004}. Most lines of sight to the GRBs have an even lower
dust-to-gas ratio than the SMC. This is especially evident for GRB
990123 for which a powerful reverse shock was observed \citep{Akerlof1999}. The only
two lines-of-sight to the GRBs with high dust-to-gas ratios are those of GRB 021004
and GRB 030329 - in both cases, the $\NH$ values are only upper limits. Yet both SEDs
are best fit by SMC dust. Incidentally, these are also the two afterglows with the
most pronounced fine structure (Paper II).}
\label{NHtoA_V}
\end{figure}

\newpage\clearpage

\begin{figure*}[ht]
\epsscale{1}
\plotone{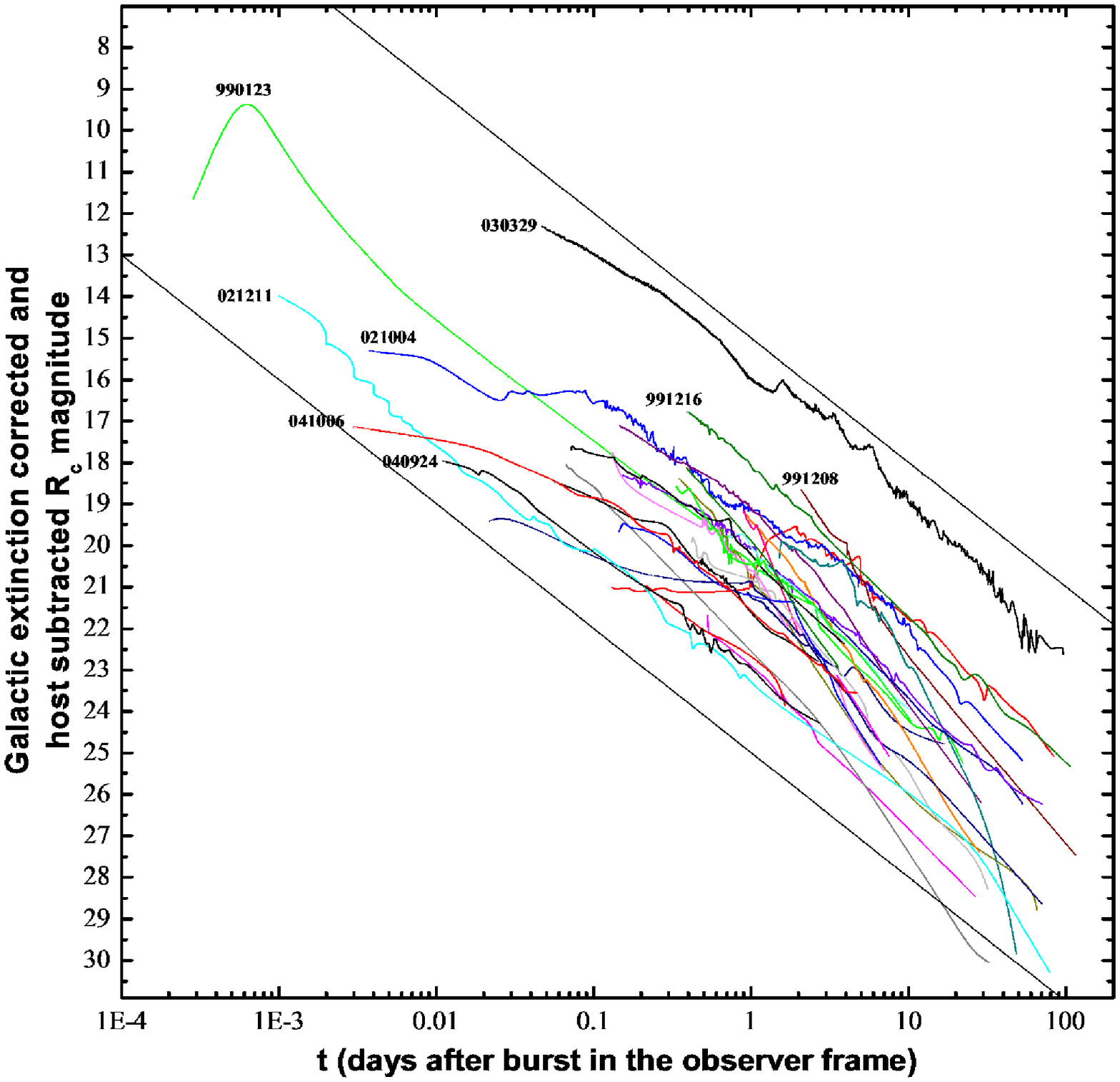}
\caption{The observed $R_C$-band light curves of all 30 afterglows in the sample of
Table \ref{tabALL}. The data have been corrected for Galactic extinction and host galaxy
contribution. In afterglows where supernova light dominates at late times, these data have
not been plotted, except for GRB 030329, where the light curve has been corrected for the
supernova contribution. At one day after the burst, there is a spread of 7.5 magnitudes,
with the afterglow of GRB 030329 being the brightest and the afterglow of GRB 021211 the
faintest. The two inclined lines are meant to guide the eye, corresponding to a decay slope
of $\alpha=1.2$.}
\label{BigPic1}
\end{figure*}

\newpage\clearpage

\begin{figure*}[ht]
\epsscale{1}
\plotone{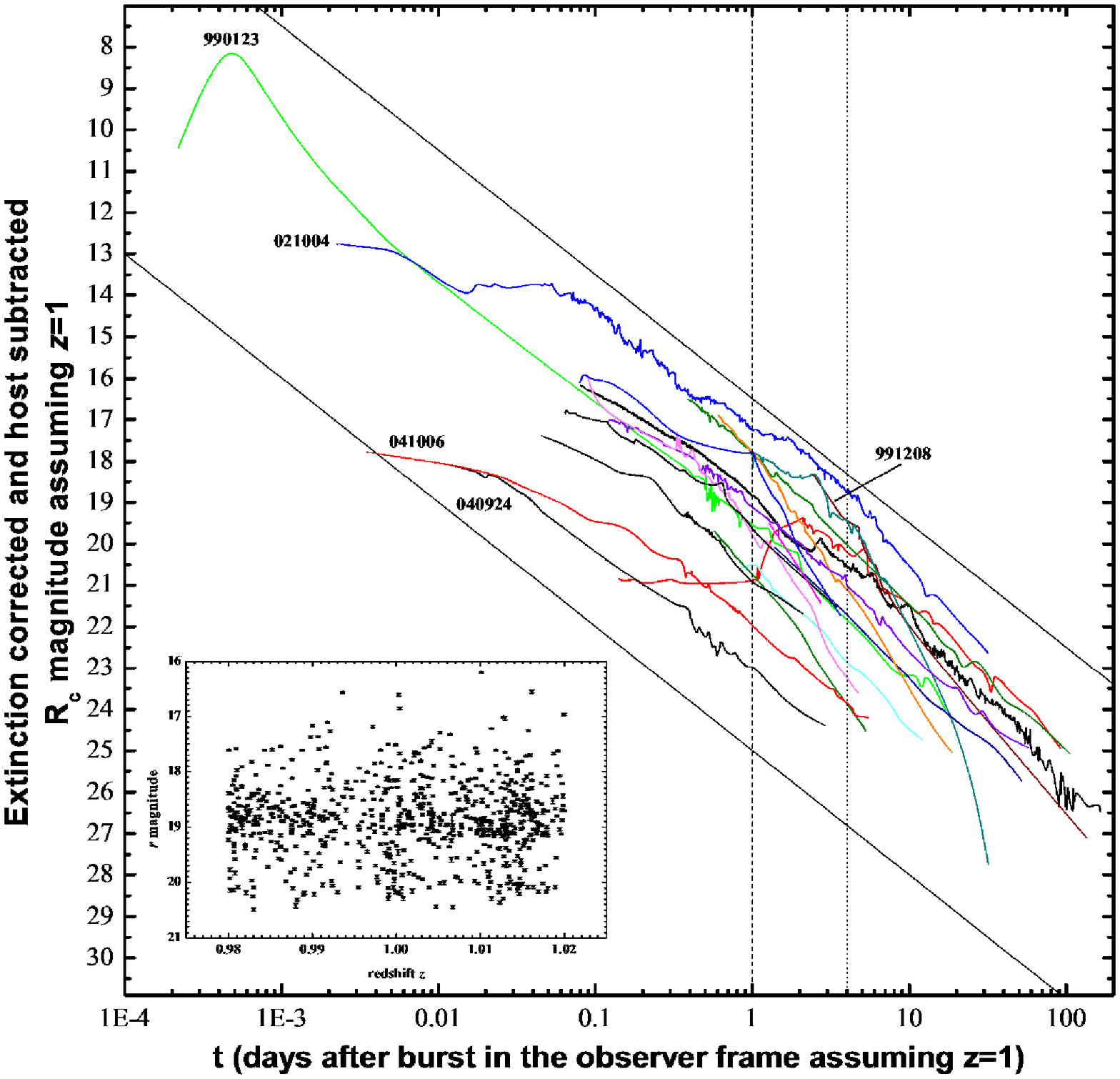}
\caption{The distribution of the apparent magnitudes of the afterglows of the Golden Sample
(Table \ref{tabGold}) after shifting them to a common redshift of $z=1$ (\kref{redshifting}).
This allows a direct comparison of the light curve evolution and luminosity. Compared with
Figure \ref{BigPic1}, the magnitude spread has been reduced. It is now 5.7 magnitudes at one
day after the burst. The brightest afterglow is GRB 021004, but it is
possible that the afterglow of GRB 991208 was even brighter at earlier times when it was not
yet discovered \citep{Castro-Tirado2001}. A large number of GRBs have afterglows of similar
brightness, these have $R_C\approx19$ at one day after the burst. Nine of the 16 afterglows
that have data at one day lie in a region only two magnitudes wide. The two inclined lines
are meant to guide the eye. The vertical lines lie at one and four days after the burst, the
times when we derive the absolute magnitude $\MB$ (Figures \ref{MB1}, \ref{MB4}). For
comparison, the inset shows 748 quasars taken from the third SDSS quasar catalog
\citep{Schneider2005} at $z\approx1$, having $\overline{r}\approx19$. 
If GRB 990123 had been at $z$=1 its optical flash would have 
peaked at $R=7.6$, which is $\approx35,000$ times as luminous.}
\label{BigPic2}
\end{figure*}

\newpage\clearpage

\begin{figure}[!t]
\epsfig{file=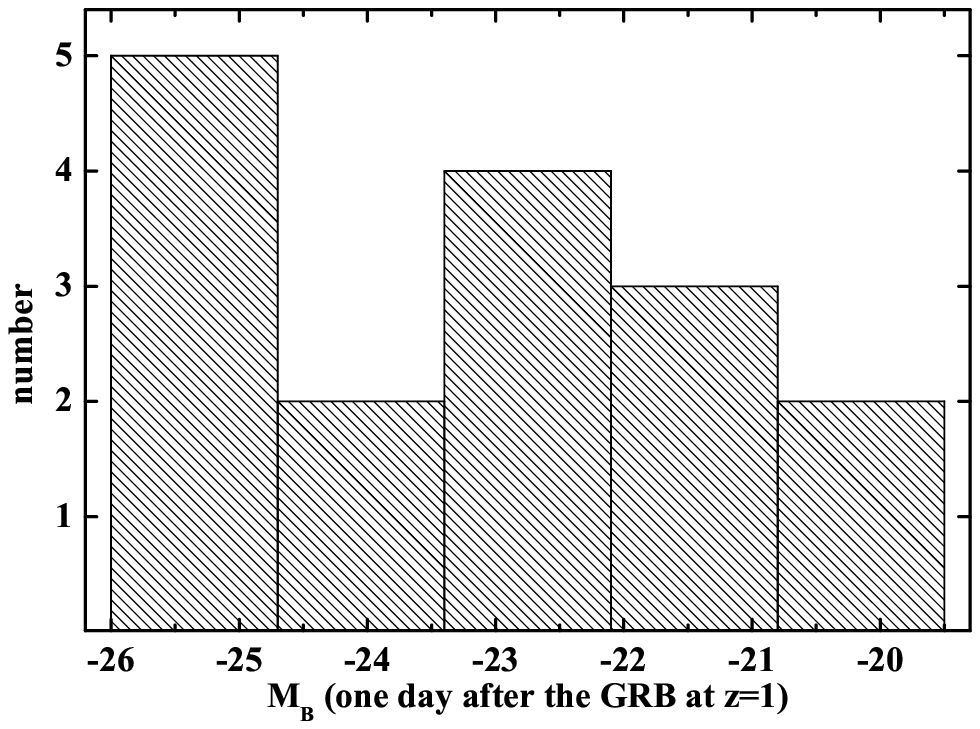,width=1\textwidth}
\caption[]{The distribution of the absolute magnitudes $\MB$,
as it follows from Fig.~\ref{BigPic2} at one day after the GRB.
The mean value is $\overline{\MB}=-23.2\pm0.4$.
Note that only 16 afterglows have observational data at that time, GRB 991208,
GRB 000131 and GRB 000911 are not included.}
\label{MB1}
\end{figure}

\newpage\clearpage

\begin{figure}[ht]
\epsfig{file=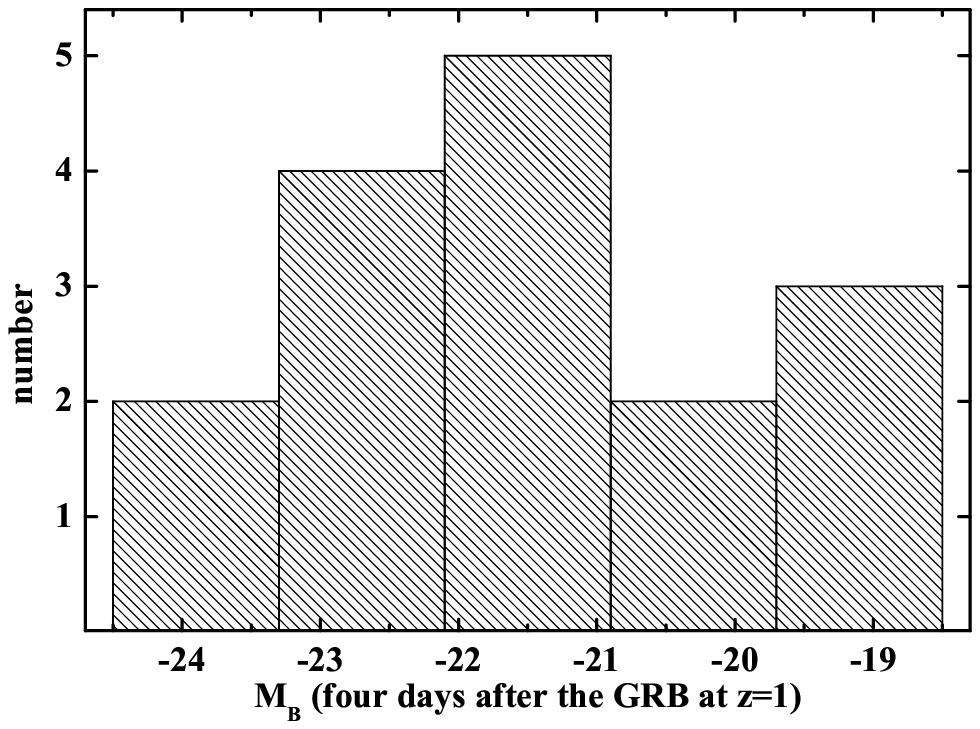,width=1\textwidth}
\caption[]{As Figure \ref{MB1}, but at four days after the GRB at $z=1$ (Table \ref{tabMB}).
The mean value is $\overline{\MB}=-21.4\pm0.4$. Note that only 16 afterglows have observational
data at that time, GRB 000131, GRB 030328 and GRB 040924 are not included.}
\label{MB4}
\end{figure}

\newpage\clearpage

\begin{figure}[ht]
\epsfig{file=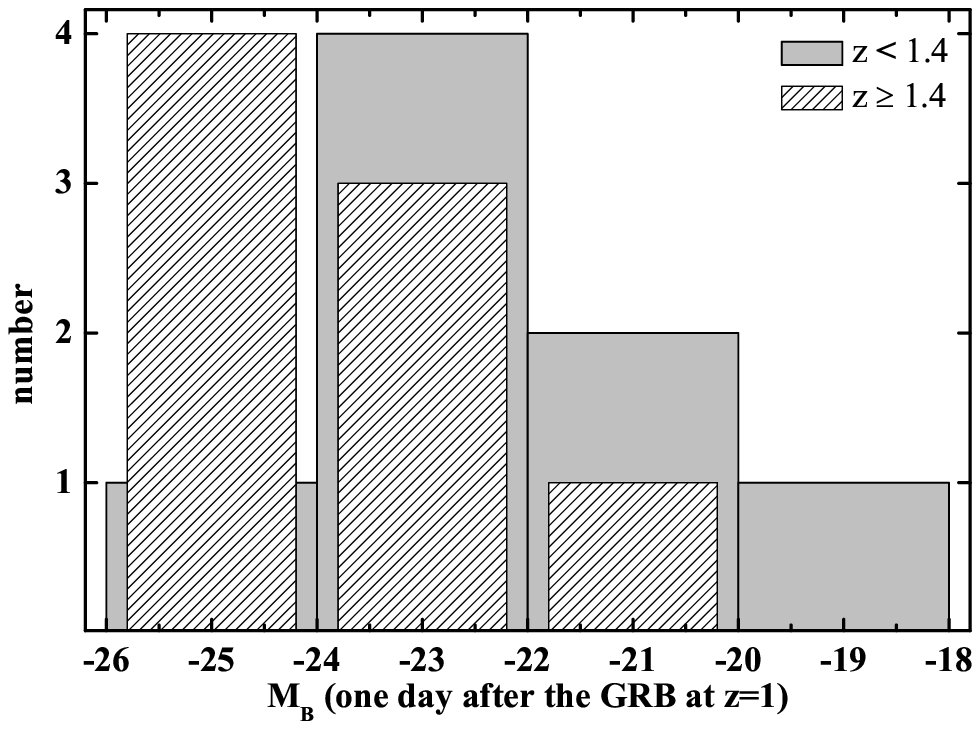,width=1\textwidth}
\caption[]{The distribution of absolute magnitudes $\MB$ at one day after the GRB at $z=1$
(Table \ref{tabMB}), divided in GRBs with $z<1.4$ and $z\geq1.4$. The means are $\overline
{\MB}=-22.4\pm0.6$ and $\overline{\MB}=-24.1\pm0.5$, respectively.}
\label{MB1z}
\end{figure}

\newpage\clearpage

\begin{figure}[ht]
\epsfig{file=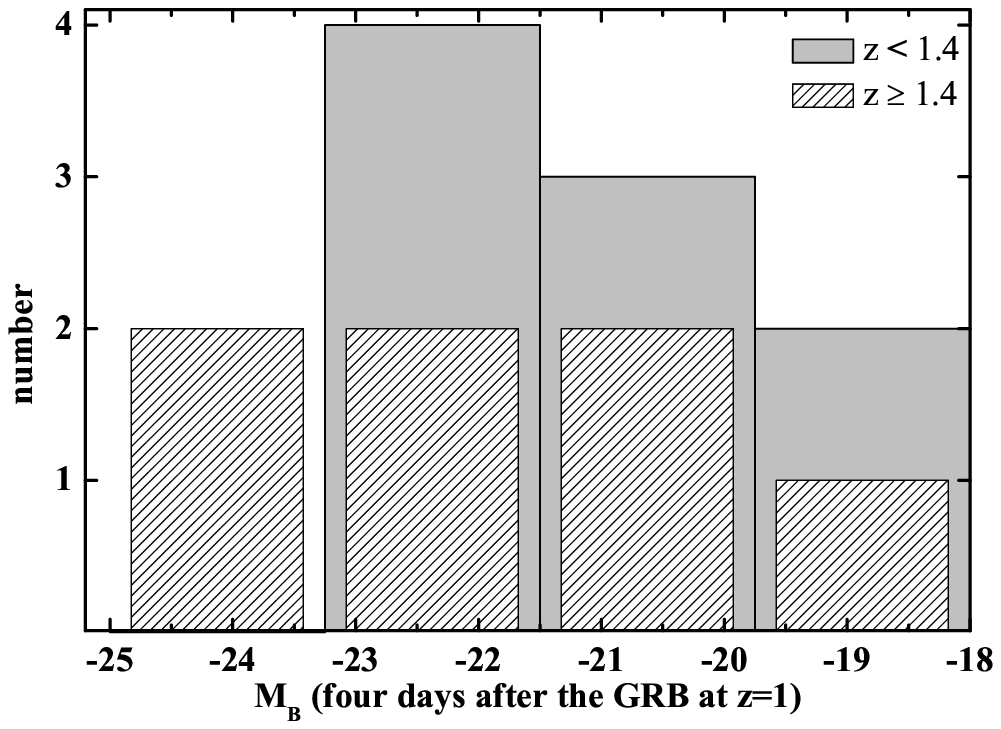,width=1\textwidth}
\caption[]{As Figure \ref{MB1z}, but at four days after the GRB (Table \ref{tabMB}). The
means are $\overline{\MB}=-21.2\pm0.5$ and $\overline{\MB}=-21.8\pm0.6$, respectively.}
\label{MB4z}
\end{figure}

\end{document}